\documentclass[10pt,twocolumn,twoside]{IEEEtran}
\pdfoutput=1
\usepackage{amsmath}
\usepackage{amsthm}
\usepackage{amssymb}
\usepackage{graphicx}

\usepackage{cite}

\ifCLASSINFOpdf
\else
\fi
\newtheorem{theorem}{Theorem}

\newcommand{\calH}{{\cal H}}
\newcommand{\expect}{\mathbb{E}}
\newcommand{\reals}{\mathbb{R}}
\newcommand{\argmax}{\operatornamewithlimits{arg\,max}}
\newcommand{\Ahat}{\widehat{A}}
\newcommand{\eps}{\varepsilon}
\newcommand{\calL}{{\cal L}}

\newcommand{\DelK}{\Delta K}

\newcommand{\khat}{\hat{k}}

\newcommand{\kmax}{k_\textrm{max}}
\newcommand{\Ghat}{\widehat{G}}
\newcommand{\ahat}{\hat{a}}

\newcommand{\xhat}{\hat{x}}
\newcommand{\Xhat}{\widehat{X}}
\newcommand{\ones}{\mathbf{1}}
\newcommand{\tr}{\operatorname{tr}}
\newcommand{\eop}{\varepsilon_1^+}
\newcommand{\eom}{\varepsilon_1^-}
\newcommand{\etp}{\varepsilon_2^+}
\newcommand{\etm}{\varepsilon_2^-}

\begin{document}
%
\title{A Spectral Framework for Anomalous Subgraph Detection}
%
%
%

\author{Benjamin~A.~Miller,~\IEEEmembership{Member,~IEEE,}
        Michelle~S.~Beard,
        Patrick~J.~Wolfe,~\IEEEmembership{Senior~Member,~IEEE,}
        and Nadya~T.~Bliss~\IEEEmembership{Senior~Member,~IEEE}
\thanks{This work is sponsored by the Assistant Secretary of Defense for Research \& Engineering under Air Force Contract FA8721-05-C-0002. Opinions, interpretations, conclusions and recommendations are those of the authors and are not necessarily endorsed by the United States Government.}%
\thanks{B. A. Miller is with Lincoln Laboratory, Massachusetts Institute of Technology, Lexington, MA, 02420 USA (e-mail: bamiller@ll.mit.edu).}
\thanks{M. S. Beard is with Charles Stark Draper Laboratory, Cambridge, MA, 02139 USA (email: mbeard@draper.com).}
\thanks{P. J. Wolfe is with the Department of Statistical Science, University College London, London WC1E 6BT UK (e-mail: p.wolfe@ucl.ac.uk).}
\thanks{N. T. Bliss is with Arizona State University, Tempe, AZ, 85287 USA (e-mail: nadya.bliss@asu.edu).}}

%
%

\markboth{In Submission to the IEEE, -- -- --~2014}%
{Miller \MakeLowercase{\textit{et al.}}: A Spectral Framework for Anomalous Subgraph Detection}
%



\maketitle

\begin{abstract}
A wide variety of application domains are concerned with data consisting of entities and their relationships or connections, formally represented as graphs. Within these diverse application areas, a common problem of interest is the detection of a subset of entities whose connectivity is anomalous with respect to the rest of the data. While the detection of such anomalous subgraphs has received a substantial amount of attention, no application-agnostic framework exists for analysis of signal detectability in graph-based data. In this paper, we describe a framework that enables such analysis using the principal eigenspace of a graph's residuals matrix, commonly called the modularity matrix in community detection. Leveraging this analytical tool, we show that the framework has a natural power metric in the spectral norm of the anomalous subgraph's adjacency matrix (signal power) and of the background graph's residuals matrix (noise power). We propose several algorithms based on spectral properties of the residuals matrix, with more computationally expensive techniques providing greater detection power. Detection and identification performance are presented for a number of signal and noise models, including clusters and bipartite foregrounds embedded into simple random backgrounds as well as graphs with community structure and realistic degree distributions. The trends observed verify intuition gleaned from other signal processing areas, such as greater detection power when the signal is embedded within a less active portion of the background. We demonstrate the utility of the proposed techniques in detecting small, highly anomalous subgraphs in real graphs derived from Internet traffic and product co-purchases.
\end{abstract}

\begin{IEEEkeywords}
Graph theory, signal detection theory, spectral analysis, residuals analysis, principal components analysis
\end{IEEEkeywords}

\ifCLASSOPTIONpeerreview
\begin{center} \bfseries EDICS Category: NET-DTIG, SSP-DETC, SSP-NGAU \end{center}
\fi
%
\IEEEpeerreviewmaketitle

\section{Introduction}
\label{sec:intro}
\IEEEPARstart{I}{n} numerous applications, the data of interest consist of entities and the relationships between them. In social network analysis, for example, the data are connections between individuals, such as who knows whom personally, who is in the same organization, or who is connected on a social networking website. In computer networks, we are often interested in which computers communicate with one another. In the natural sciences, we may want to know which chemicals interact in a reaction. Across these varied domains, data regarding connections, relationships and interactions between discrete entities enhances situational awareness and diversifies by incorporating additional contextual information.

When working with relational data, it is common to formally represent the relationships as a graph. A graph $G=(V,E)$ is a pair of sets: a set of vertices, $V$, comprising the entities, and a set of edges, $E$, denoting relationships between them. Graph theory provides an abstract mathematical object that has been applied in all of the above contexts. Indeed, graphs have been used to model protein interactions \cite{yeast} and to represent communication between computers \cite{eigenspaceAnomalyDetection}. Graphs---commonly called networks in practice---are used extensively in social network analysis, with many graph algorithms focused on detection of communities \cite{newman06,Xu14b} and influential figures \cite{authoritativeSources}.

As a data structure, graphs have long been utilized by signal processing practitioners. Analysis of graphs derived from radio frequency or image data is common, as a graph structure can help classify similar measurements (see, e.g., \cite{changeDetectionSAR}). Recent research has also defined traditional digital signal processing kernels---such as filtering and Fourier transforms---for signals that propagate along edges in a graph \cite{DSPonGraphs, SPoG}. When the graph comprises the data itself, rather than a means of storage, significant complications arise. Graphs are discrete, combinatorial structures, and, thus, they lack the convenient mathematical context of Euclidean vector spaces. The ability to perform linear transformations and the analytical tractability of working with Gaussian noise are not available in general when working with relational data. Deriving an optimal detector for a small signal subgraph buried within a large network, then, becomes potentially intractable, as it may require the solution to an NP-hard problem.

Despite these complications, it is desirable to understand notions of detectability of small subgraphs embedded within a large background. The ability to detect small signals in these contexts would be useful in many domains, from the detection of malicious traffic in a computer network to the discovery of threatening activity in a social network. Recent work in this area has considered subgraph detection from a variety of perspectives. Work has been done on detection of specific target subgraphs in random backgrounds \cite{randomTerroristTransactions}, with special attention paid in the computer science and statistics communities to planted cliques \cite{Alon1998, NadakuditiPlantedClique} and planted clusters \cite{AriasCastro13, ACSparse2013}. Other work assumes common substructures over the graph, and detects anomalies based on deviations from the ``normative pattern'' via methods such as minimum description length \cite{anomDet07} or analysis of the graph Laplacian \cite{Skillicorn07}. Techniques such as threat propagation \cite{SmithICASSP2012,SmithNetworkDetection2}
 and vertex nomination \cite{Coppersmith12} consider a cue vertex as a knowledge prior, giving an initial indication of which vertices are of interest, the objective then being to find the remainder of the subgraph. Community detection in graphs is a widely studied related problem \cite{FortunatoSurvey}, where the communities in the graph are sometimes cast as deviations from a null hypothesis in which the graph has no community structure \cite{NewmanGirvan04}. 

The objective of the present contribution is to develop a broadly applicable detection framework for graph-based data. To apply in these varied domains, this framework should be independent of the specific application. We focus specifically on the uncued anomalous subgraph detection problem, where the goal is to detect the presence of a subgraph that is a statistical outlier without a ``tip'' vertex provided as a cue. As graphs of interest are often extremely large, the framework should have favorable scaling properties as the number of vertices and edges grow. To gain insight into properties that influence subgraph detectability, the framework will ideally have a natural metric for signal and noise power, to enable discussion of quantities like signal-to-noise ratio that are intrinsic to signal processing applications.

In this paper, we present a spectral framework to address the uncued subgraph detection. 
 This framework is based on a regression-style analysis of residuals, in which an observed random graph is compared to its expected value to find outliers. We analyze the graph in the space of the principal eigenvectors of its residuals matrix, which offers two advantages: it allows us to use results from spectral graph theory to elucidate the notion of subgraph detectability, and works within in a linear algebraic framework with which many signal processing researchers are familiar. Within this framework, the spectral norm provides a good metric for signal and noise power, as we demonstrate analytically and empirically. This framework also enables the development of algorithms that work in a low-dimensional space to detect small anomalies, several of which are discussed in this paper. 

The remainder of this paper is organized as follows. 
In Section \ref{sec:model}, we formally define the subgraph detection problem. 
Section \ref{sec:related} provides a brief summary of related work on subgraph detection and graph residuals analysis. 
Section \ref{sec:framework} details our proposed subgraph detection framework. In Section \ref{sec:algorithms}, we outline several algorithms for anomaly detection within the framework. Section \ref{sec:results} presents detection results for several simulated datasets, and in Section \ref{sec:applications} we demonstrate these techniques on real datasets. Finally, in Section \ref{sec:conclusion}, we summarize and discuss open problems and ongoing work.

\section{Problem Model}
\label{sec:model}
\subsection{Definitions and Notation}
\label{subsec:def}
In the subgraph detection problem, the observation is a graph $G=(V,E)$. We will denote the sizes of the vertex and edge sets as $N=|V|$ and $M=|E|$, respectively. A subgraph $G_S=(V_S, E_S)$ of $G$ is a graph in which $V_S\subset V$ and $E_S\subset E\cap (V_S\times V_S)$, where the Cartesian product $V\times V$ is the set of all possible edges in a graph with vertex set $V$. 
In this paper, we consider graphs whose edges are unweighted and undirected. We will allow the possibility of self-loops, meaning an edge may connect vertex to itself. Since edges have no weight, two graphs will be combined via their union. The union of two graphs, $G_1=(V_1,E_1)$ and $G_2=(V_2,E_2)$, is defined as $G_1 \cup G_2 = (V_1\cup V_2, E_1 \cup E_2).$

Working in a spectral framework, we will make use of matrix representations for graphs. The adjacency matrix $A=\{a_{ij}\}$ of $G$ is a binary $N\times N$ matrix. Each row and column is associated with a vertex in $V$. This implies an arbitrary ordering of the vertices with integers from $1$ to $N$, and we will denote the $i$th vertex $v_i$. Then $a_{ij}$ is 1 if there is an edge connecting $v_i$ and $v_j$, and is 0 otherwise. Similarly, let $A_S=\{s_{ij}\}$ be the adjacency matrix for the signal subgraph. Since we consider undirected graphs, $A$ and $A_S$ are symmetric. Matrix norms will also be used in the discussion of signal and noise power. Unless otherwise noted, the matrix norm will be the spectral norm, i.e., the induced $L_2$ norm,
\begin{equation}
\|A\|=\max_{\|x\|_2=1}{\|Ax\|_2},
\label{eqn:matrixNorm}
\end{equation}
which is equivalent to the absolute value of the largest-magnitude eigenvalue of the matrix.

Our framework is focused on detection of signals within a random background. 
The analysis presented in this paper is based on the assumption of Bernoulli random graphs, where 
the probability of an edge between $v_i$ and $v_j$ is a Bernoulli random variable with expected value $p_{ij}$. Note that the edge probabilities may be different for all pairs of vertices. Since the presence of each edge is a Bernoulli random variable, the expected value of $A$ is given by $P=\{p_{ij}\}$. We refer to $P$ as the probability matrix of the graph.

Another important notion when dealing with graphs is degree. A vertex's degree is the number of edges adjacent to the vertex. The observed degree of vertex $v_i$ will be denoted $k_i$, and its expected degree is denoted $\expect\left[k_i\right]=d_i$. Note that $k_i=\sum_{j=1}^{N}{a_{ij}}$ and $d_i=\sum_{j=1}^{N}{p_{ij}}$\footnote{Using this convention, a self-loop only increases a vertex's degree by 1.}. 
The vectors of the observed and expected degrees will be denoted $k$ and $d$, respectively. The volume of the graph, $\operatorname{Vol}(G),$ is the sum of the degrees over all vertices.

\subsection{The Subgraph Detection Problem}
In some cases, the observed graph $G$ will consist of only typical background activity. This is the ``noise only'' scenario. In other cases, most of $G$ exhibits typical behavior, but a small subgraph has an anomalous topology. This is the ``signal-plus-noise'' scenario. In this case, the noise graph, denoted $G_N=(V_N, E_N)$, and the signal subgraph, $G_S=(V_S, E_S)$ are combined via union.

The objective, given the observation $G$, is to discriminate between the two scenarios. Formally, we want to resolve the following binary hypothesis test:
\begin{align}
\label{eqn:HypTest}
\begin{cases}
\calH_{0}: & G = G_N \\
\calH_{1}: &  G = G_N \cup G_{S}.
\end{cases}
\end{align}
Thus, we have the classical signal detection problem: under the null hypothesis $\calH_0$, the observation is purely noise, while under the alternative hypothesis $\calH_1$, a signal is also present. Here $G_N$ and $G_S$ are both random graphs, with $G_N$ drawn from the noise distribution and $G_S$ drawn from the signal distribution. We will only consider cases in which the vertex set of the signal subgraph is a subset of the vertices in the background, i.e., $V_S\subset V_N=V$.

\section{Related Work}
\label{sec:related}
While there are many flavors of subgraph detection research, not all of them work under the same assumptions as in this paper. For example, we consider a variety of noise models, which may not have the ``normative pattern'' required to use techniques based on common subgraphs \cite{anomDet07, Skillicorn07}. 
 Research into anomaly detection in dynamic graphs by Priebe et al. \cite{scanStats} uses the history of a node's neighborhood to detect anomalous behavior, but this would not apply in the case of static graphs, which is the focus of this work. As our interest is in uncued techniques, we operate in a different context from the work in \cite{SmithICASSP2012,SmithNetworkDetection2,Coppersmith12}. These methods are complementary to the techniques outlined in this paper, as a set of outlier vertices could be used to seed a cued algorithm and do further exploration.


Previous work has considered optimal detection in the same context we consider in this paper, though in a restricted setting. In \cite{randomTerroristTransactions}, the authors consider the detection of a specific foreground embedded (via union) into a large graph in which each possible edge occurs with equal probability (i.e., the random graph model of Erd\H{o}s and R\'{e}nyi). In this setting, the likelihood ratio can be written in closed form, as demonstrated by the following theorem.

\begin{theorem}[Mifflin et al. \cite{randomTerroristTransactions}]
Let $G$ denote the random graph where each possible edge occurs with equal probability $p$, and let $H$ denote the target graph. The likelihood ratio of an observed graph $J$ is 
\begin{equation}
\Lambda_H(J)=\frac{X_H(J)}{\expect\left[X_H(G)\right]}.
\end{equation}
\label{thm:Mifflin}
\end{theorem}\noindent
Here $X_H(\cdot)$ denotes the number of occurrences of $H$ in the graph. The applicability of this result, therefore, requires a tractable way to count all subgraphs of the observation $J$ that are isomorphic with the target. This is NP-hard in general \cite{CLR}, although 
there may be feasible methods to accomplish this for certain targets within sparse backgrounds.

While the previous example requires a complicated procedure, detection of random subgraphs embedded into random backgrounds may be an even harder problem. Take, for example, the detection problem where the background and foreground are both Erd\H{o}s--R\'{e}nyi, i.e., when the null and alternative hypotheses are given by
\begin{align}
\begin{cases}
\calH_{0}: & \textrm{each pair of vertices shares an edge with}\\
           & \textrm{probability $p$} \\
\calH_{1}: & \textrm{an $N_S$-vertex subgraph was embedded whose}\\
           & \textrm{ edges were generated with probability $p_S$}.
\end{cases}
\label{eqn:ERinER}
\end{align}
In this situation, we can derive an optimal detection statistic.
\begin{theorem}
For an observed graph $G=(V,E)$, let $X$ be a subset of $V$ of size $N_S$, and $E_X\subset E$ be the set of all edges existing between the vertices in $X$. The likelihood ratio for resolving the hypothesis test in (\ref{eqn:ERinER}) is given by\begin{equation}
\binom{N}{N_S}^{-1}\left(\frac{1-\hat{p}}{1-p}\right)^{\binom{N_S}{2}}\sum_{\begin{subarray}{c} X\subset V\\ |X|=N_S \end{subarray}}{\left[\frac{\hat{p}(1-p)}{p(1-\hat{p})}\right]^{|E_X|}},
\label{eqn:ERinERlikelihood}
\end{equation}
where $\hat{p}=p+p_S-pp_S$.
\label{thm:ERinER}
\end{theorem}\noindent
A proof of Theorem \ref{thm:ERinER} is provided in Appendix \ref{app:likelihoodProof}. Even in this relatively simple scenario, computing the likelihood ratio in (\ref{eqn:ERinERlikelihood}) requires, at least, knowing how many $N_S$-vertex induced subgraphs contain each possible number of edges. In \cite{AriasCastro13}, it is shown that some computable tests asymptotically achieve the information-theoretic bound for dense backgrounds, but there are no known polynomial-time algorithms that achieve the bound in a sparse graph \cite{ACSparse2013}. For more complicated models, calculating the optimal detection statistic is likely to be even more difficult.

The subgraph detection framework presented in this paper is based on graph residuals analysis. The residuals of a random graph are the difference between the observed graph and its expected value\footnote{This is distinct, it should be noted, from the notion of residual networks when computing network flow \cite{CLR}.}. For a random graph $G$, we analyze its residuals matrix
\begin{equation}
B:=A-\expect\left[A\right].
\label{eqn:residuals}
\end{equation}
In the area of community detection, a widely used quantity to evaluate the quality of separation of a graph into communities is modularity, defined in \cite{NewmanGirvan04}. The modularity of a partition $C=\{C_1,\cdots, C_n\}$ is defined as
\begin{equation}
Q=\sum_{i=1}^{n}{(e_{ii}-a_i^2)},
\label{eqn:modularity}
\end{equation}
where $C_i$ are disjoint subsets of $V$ covering the entire set, $e_{ii}$ is the proportion of edges entirely within $C_i$, and $a_i$ is the proportion of edge connections in $C_i$, i.e.,
\begin{equation}
a_i=\sum_{j=1}^{n}{e_{ij}},
\end{equation}
with $e_{ij}$ denoting half the number of edges between $C_i$ and $C_j$ for $i\neq j$ (half to prevent from counting the edge in both $e_{ij}$ and $e_{ji}$). Note that $a_i^2$ is the expected proportion of edges within $C_i$ if the edges were randomly rewired (i.e., the degree of each vertex is preserved, but edges are cut and reconnected at random). Indeed, if the edge proportions are the only thing maintained in the rewiring, the fraction of edges from any community that connect to a vertex in $C_i$ will be $a_i$. Thus, the proportion of the total edges from $C_i$ to $C_j$ will be $a_ia_j$. Taken as an analysis of deviations from an expected topology, modularity is a residuals-based quantity.

In the community detection literature, numerous algorithms exist to maximize $Q$ for a given number of communities. In \cite{newman06}, an algorithm is proposed by casting modularity maximization as optimization of a vector with respect to a matrix. The modularity matrix $B$ is given as the observed minus the expected adjacency matrices, i.e., a matrix of the form in (\ref{eqn:residuals}). To divide the graph into two partitions in which modularity is maximized, we can solve
\begin{equation}
\hat{s}=\argmax_{s\in \{-1,1\}^N}{s^T\left(A-\frac{1}{\operatorname{Vol}(G)}kk^T\right)s},
\label{eqn:modMax}
\end{equation}
and declare the vertices corresponding to the positive entries of $\hat{s}$ to be in one community, with the negative entries indicating the other. This technique will optimize $Q$ for a partition into 2 communities. Since this is a hard problem, it is suggested that the principal eigenvector of
\begin{equation}
B= A-\frac{1}{\operatorname{Vol}(G)}kk^T
\label{eqn:standardModMat}
\end{equation}
is computed---thereby relaxing the problem into the real numbers---with the same strategy of discriminating based on the sign of eigenvector components used to divide the graph into communities.

This is an example of a community detection algorithm based on spectral properties of a graph, which have inspired a significant amount of work in the detection of communities \cite{newman06, RuanZhangSpectral, WhiteSmythClustering, algebraicModularity} and global anomalies \cite{eigenspaceAnomalyDetection, DingKolaczyk13, anomEigenCompression}. 
In this paper, we leverage these same properties within a novel framework for detection of small subgraphs whose behavior is distinct from background activity.

\section{Detection Framework}
\label{sec:framework}
\subsection{Framework Overview}
\label{subsec:overview}
The subgraph detection framework we propose is based on the analysis of graph residuals, as expressed by (\ref{eqn:residuals}). We may be given $\expect[A]$, or it may be estimated from the observed data. This is similar to analysis of variance in linear regression: We compare the observed data to its expectation, and if the deviations from the expected value are not consistent with variations due to noise, then this may indicate the presence of a signal (in this case an anomalous subgraph).

To reduce the dimensionality of the problem, this framework deals with a graph's spectral properties. Using the principal components of the residuals matrix, we can consider a graph in the linear subspace in which its residuals are largest. For some established models, there is also theory regarding the eigenvalues and eigenvectors of these matrices \cite{ChungSGTBook}. This technique is used in community detection, and is similar to models in which each vertex has a position in a latent Euclidean space (see, e.g., \cite{RandomDotProduct}). 
The presence of certain anomalous subgraphs will alter the projection of a graph into this Euclidean residuals space. Working within this space, we can compute test statistics and, from these, resolve the hypothesis test (\ref{eqn:HypTest}). While these will not be optimal detection statistics as in Theorems \ref{thm:Mifflin} and \ref{thm:ERinER}, this framework can be applied to a wide variety of random graph models, is computationally tractable, and, as we demonstrate in subsequent sections, is quite useful for resolving the subgraph detection problem in a variety of scenarios.

We use the modularity matrix from (\ref{eqn:modMax}) as our baseline residuals model. This has several advantages. First, the ``given expected degree'' model has been well studied, and we know properties of its eigenvalues and eigenvectors \cite{RandomGraphSpectra}. Second, the model's expected value term is low-rank, which allows easy computation of the eigenvectors of $B$ without computing a dense $N\times N$ matrix (as noted in \cite{newman06} and described in \cite{ICASSP2012}). This makes the model computationally tractable for large graphs where algorithms more expensive than $O(M)$ can be prohibitive. This model also has a simple fitting procedure. Estimating the expected degree as simply the observed degree is, in fact, the maximum likelihood estimator for the version of this model where each possible edge is a Poisson random variable \cite{PerryWolfe12}. For small edge probabilities, this is a good approximation for Bernoulli random variables. Finally, this model has demonstrated utility for inter-community behavior; i.e., the probability of connections between vertices in different communities seems to follow such a model (the reason that observed degree was added as a covariate in \cite{ChoiWolfeAiroldi11}).

\subsection{Power Metrics}
As mentioned previously, one important aspect of a signal processing framework is a metric for signal and noise power. This provides a quantity that enables an intuitive assessment of the detectability of a signal in a given background. Again, vector signals with Gaussian noise provide an intuitive metric based on vector norms, while such quantities are less clear in the context of random graphs.

There are several intuitive quantities that could be used for signal or noise power in the context of random graphs. One natural choice would be number of edges, or perhaps average degree. It seems intuitive that a signal graph with a large number of edges would be easier to detect, and that greater variance in the number of edges in the background would make this more difficult. A related linear algebraic quantity would be the Frobenius norm of the residuals matrix, i.e., the sum of the squared residuals over all ordered pairs of vertices. This would consider each edge probability separately, emphasizing the presence of less-likely edges.

These metrics, however, have a few shortcomings. In both cases, the signal power measurement will be exactly the same for any subgraph with the same number of edges. Consider two different trees: a path, in which each edge can be traversed while visiting each vertex exactly once; and a star, where one vertex is connected to all others. Both will have $N_S-1$ edges, and a Frobenius norm of $2(N_S-1)$. The star, however, is much more concentrated on one vertex, and this will cause it to stand out more in the eigenspace (it is also much less likely to occur by chance if edges are randomly placed). The power metric we use should provide an indication of a subgraph's likelihood to stand apart from the background in the eigenspace, since this is the space in which we consider the data.

Working within a spectral framework, the spectral norm defined in (\ref{eqn:matrixNorm}) provides a natural power metric. Using $\|A-\expect[A]\|$ as a metric for noise power, and $\|A_S\|$ as a metric for signal power, we can determine the detectability of a subgraph in principal eigenspace. To see this, we first define a new matrix, $\Ahat=\{\ahat_{ij}\}$, which is the adjacency matrix of $\Ghat=(V, E_S\setminus E)$, i.e., the edges of the anomaly that do not appear in the background. For deterministic foreground graphs, if $s_{ij}$ is 1, then $\ahat_{ij}$ is a random variable whose value is 1 with probability $1-p_{ij}$ and 0 with probability $p_{ij}$. For a random Bernoulli foreground, if $\expect[s_{ij}]=q_{ij}$, then $\ahat_{ij}$ is 1 with probability $q_{ij}(1-p_{ij})$. Thus, when the subgraph is embedded within vertices where the interaction level is low, $\expect[\Ahat]\approx\expect[A_S]$. For convenience, we will also denote a partition of the residuals matrix as 
\begin{equation}
B=\left[
\begin{array}{cc}
B_S & B_{SN}\\
B_{SN}^T & B_N
\end{array}\right],
\end{equation}
where the rows and columns have been permuted so that the subgraph vertices are those with the smallest indices. The submatrix $B_S$ is the background residuals within the subgraph vertices, $B_{SN}$ is the residuals occurring between the subgraph and the rest of the graph, and $B_N$ includes only the residuals within the complement of the subgraph vertices.

If the spectral norm of the signal subgraph is sufficiently large with respect to the background power, the subgraph will dominate the principal eigenvector of the residuals matrix. This is captured in the following theorem, a proof of which is provided in Appendix \ref{app:SNProof}.
\begin{theorem}
\label{thm:SN}
Let $B$ be the residuals matrix of a graph drawn from an arbitrary Bernoulli graph process, and $\Ahat$ be the adjacency matrix of the subgraph that does not include edges in the background graph. If $u$ is the unit eigenvector associated with the largest positive eigenvalue of $B+\Ahat$ (the residuals matrix after embedding), then assuming $\|\Ahat\|>\|B_N\|+\|B_S\|$, the components of $u$ associated with only the signal vertices, denoted $u_S$, is bounded below as $\|u_S\|_2^2\geq1-\eps$ where 
\begin{equation}
\eps=O\left(\frac{\left(\|\Ahat\|-\|B_N\|\right)\left(\|B_S\|+\|B_{SN}\|\right)+\|B_{SN}\|^2}{\left(\|\Ahat\|-\|B_N\|\right)^2+\|B_{SN}\|^2}\right).
\label{eqn:normBound}
\end{equation}
\end{theorem}

Consider the implication of Theorem \ref{thm:SN} for a fixed background, when embedding on a fixed subset of vertices. The theorem states that as the difference between the signal power and the power of the noise among the non-signal vertices ($\|\Ahat\|-\|B_N\|$) becomes much larger than the noise power involving subgraph vertices ($\|B_S\|+\|B_{SN}\|$), the principal eigenvector will become concentrated on the foreground vertices. 
 A few aspects of this theorem confirm intuition from other signal processing areas. First, if there is significant noise activity within the subgraph vertices, then $\|\Ahat\|$ may be significantly smaller than $\|A_S\|$, and $\|B_S\|$ may be relatively large. This means that a signal placed in strong noise will be difficult to detect, which is always the case in detection problems. Also, the bound in the theorem shows that if a relatively strong subgraph is embedded where there is typically very little activity, and where there is relatively little interaction with the remainder of the graph (i.e., small $\|B_S\|$ and $\|B_{SN}\|$), the subgraph will be much easier to detect. Put in traditional signal processing language, the signal will be much easier to detect when it is less correlated with the noise. Working within this framework, we see the same properties of the interaction between signal and noise that affect detectability in domains like radar and communications.

\begin{figure}
\begin{centering}
\includegraphics[width=2.3in]{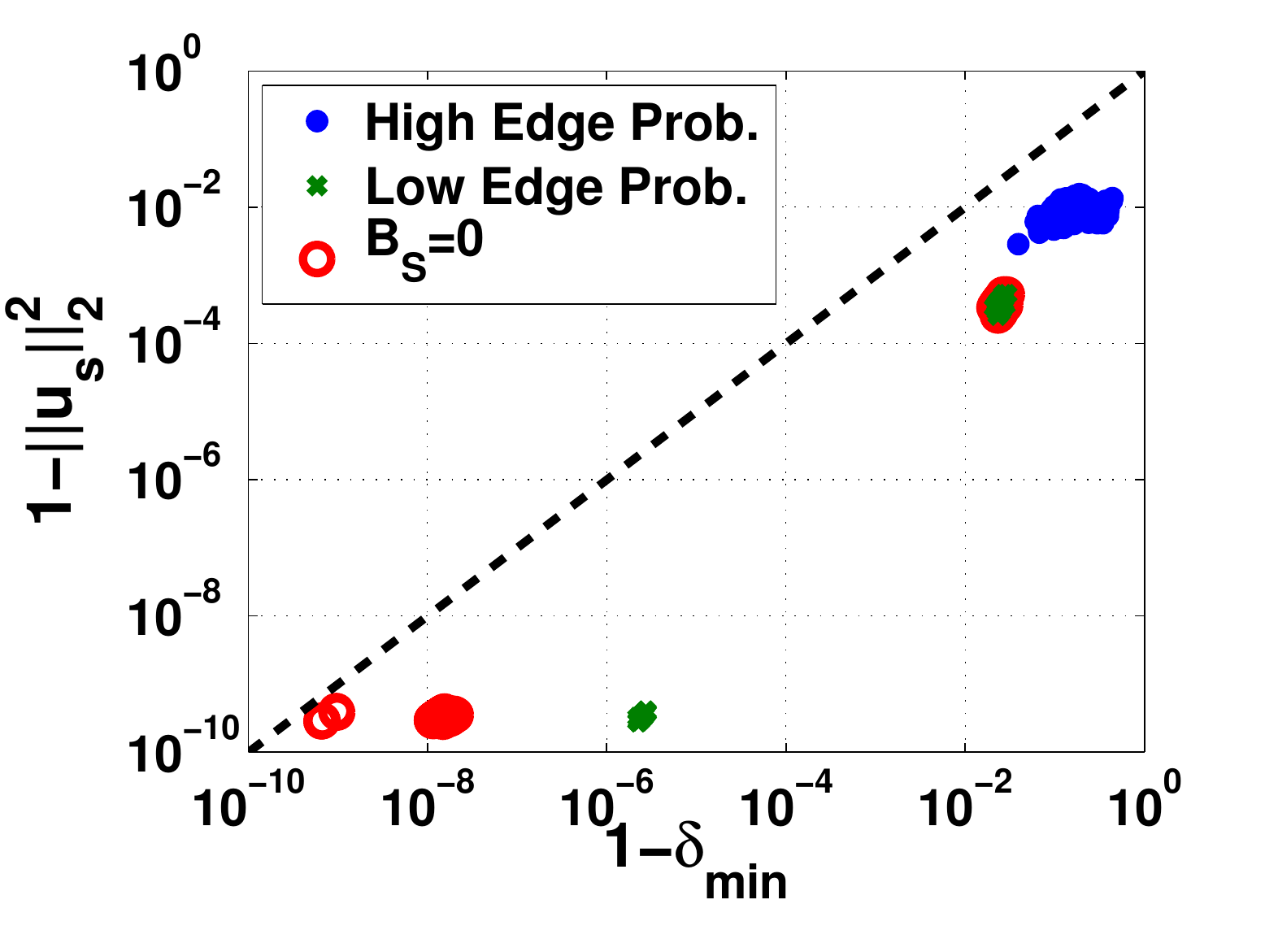}
\caption{Empirical comparison to bound in Theorem \ref{thm:SN}. The bound holds for each case in this scenario with a 4096-vertex random background and a 15-vertex dense signal subgraph, though it is only tight for cases where $B_S=0$.}
\label{fig:normBound}
\end{centering}
\end{figure}

An empirical example is provided in Fig.~\ref{fig:normBound}. In this case, a 4096-vertex Erd\H{o}s--R\'{e}nyi graph (see Section \ref{subsubsection:erdosRenyi}) is generated, with a 15-vertex subgraph with 90\% edge probability embedded. The horizontal axis is $1-\delta_{\min}$, where $\delta_{\min}$ is the expression in (\ref{eqn:LB}) in Appendix \ref{app:SNProof}. The bound holds for all cases considered, and the empirical results often are an order of magnitude below the maximum for both the higher and lower edge probabilities ($p=4\times10^{-4}$ and $p=1\times10^{-6}$, respectively). Only when a case is considered where there is no background connectivity within the subgraph vertices is the bound approached more closely.

\section{Detection Algorithms}
\label{sec:algorithms}
For relatively large subgraph anomalies, a simple ``energy detector'' based on the spectral norm of the residuals matrix will provide good detection performance. It is desirable, however, to be able to detect much smaller subgraphs, which may not stand out in the principal eigenvector. A few techniques have been developed within this framework to detect subtler anomalies \cite{ICASSP2010, NIPS2010, SSP2011_Navraj, SPGLLJ}, which we outline in this section.
\subsection{Chi-Squared Statistic in Principal Components}
\label{subsec:chi2}
\begin{figure}
\begin{center}
\includegraphics[width=1.6in]{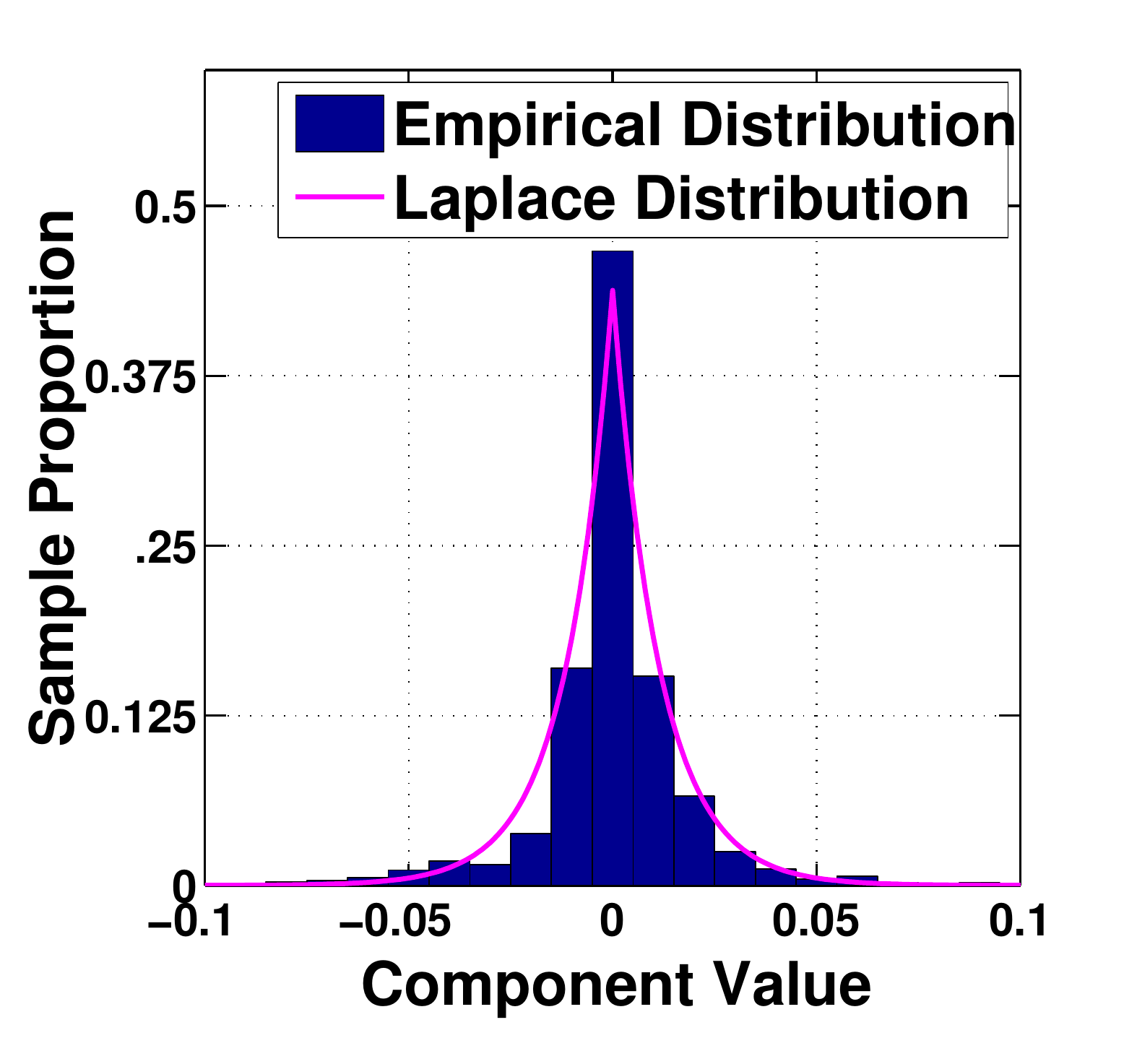}
\includegraphics[width=1.6in]{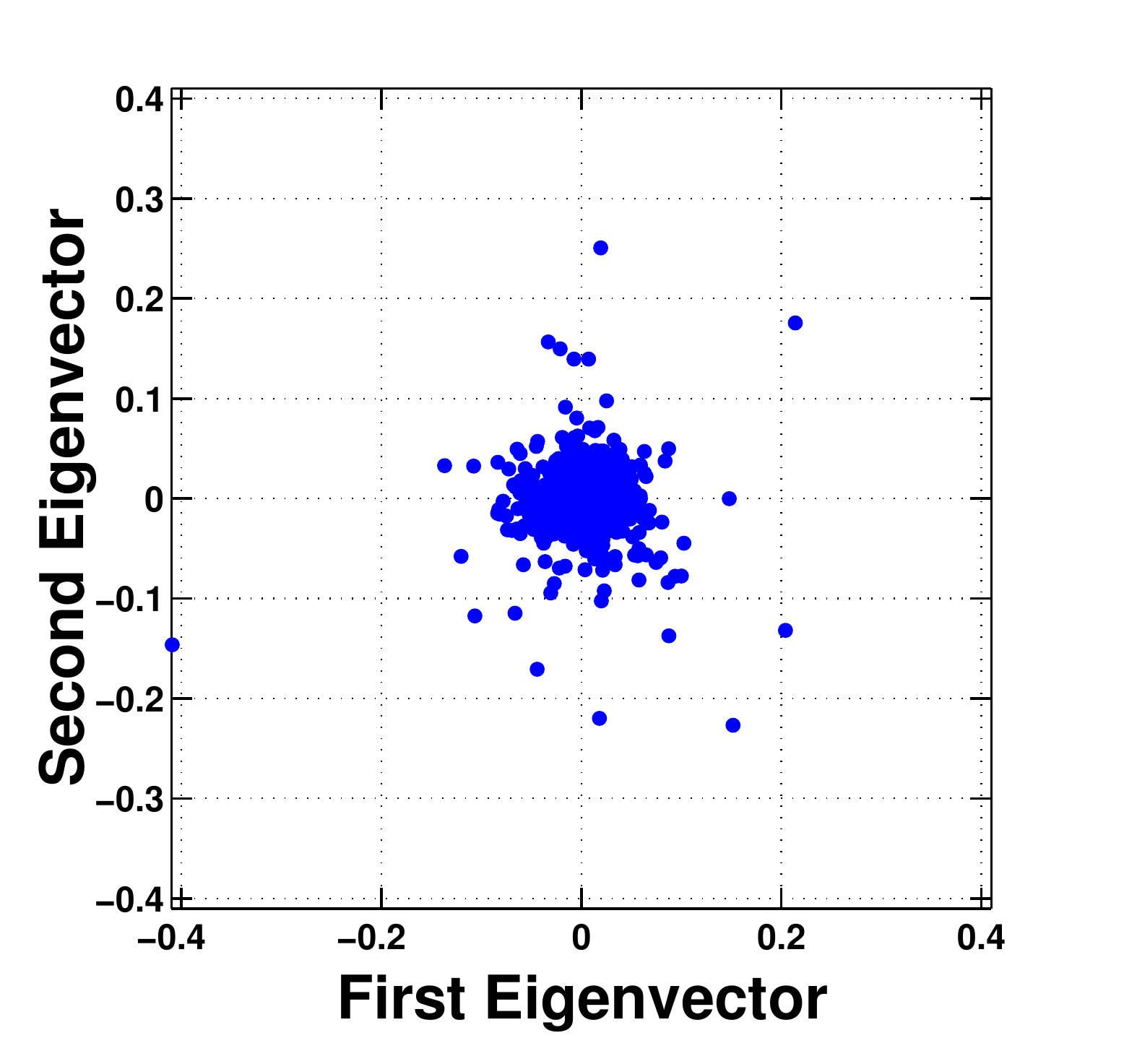}
\caption{Distributions of vertex components in principal eigenvectors: a histogram of components in the first eigenvector (left), with a comparison to a Laplace distribution, and a scatterplot (right) in the principal 2-dimensional subspace, demonstrating its radial symmetry.}
\label{fig:eigenDist}
\end{center}
\end{figure}

The first algorithm is based on the symmetry of the projection of $B$ into its 2 principal components. This will enable the detection of subgraphs that do not stand out in the first eigenvector. We have empirically observed for several random graph models that, when projecting the residuals into their principal two components, the result is rather radially symmetric. For sparse graphs, the entries in the principal eigenvectors resemble a Laplace distribution, as shown on the left in Fig. \ref{fig:eigenDist}, which is consistent with behavior observed in sparse Erd\H{o}s--R\'{e}nyi graphs. 
The righthand plot in Fig. \ref{fig:eigenDist} demonstrates the symmetry of the residuals in the top two eigenvectors.

When an anomaly is embedded within the graph, as previously discussed, the subgraph vertices will stand apart from the background. Therefore, we compute a statistic that is based on symmetry in this space to detect the presence of an anomaly. The detection statistic is a chi-squared statistic based on a $2\times 2$ contingency table, where the table contains the number of vertices projected into each quadrant of the two-dimensional space. (That is, the number of rows of $[u_1, u_2]$, where $u_1$ and $u_2$ are (column) eigenvectors of $B$, fall into each quadrant.) This yields a $2\times 2$ matrix $O=\{o_{ij}\}$ of the observed numbers of points in each section. From the observation, we compute the expected number of points under the assumption of independence, $\bar{O}=\{\bar{o}_{ij}\}$, where
\begin{equation}
\bar{o}_{ij}=\frac{(o_{i1}+o_{i2})(o_{1j}+o_{2j})}{N}.
\end{equation}
The chi-squared statistic is then calculated as
\begin{equation}
\chi^2([u_1 u_2])=\sum_i\sum_j{\frac{(o_{ij}-\bar{o}_{ij})^2}{\bar{o}_{ij}}},
\end{equation}
and, to favor radial symmetry, we maximize the statistic over rotation in the plane, computing
\begin{equation}
\chi_{\max}^2=\max_{\theta}{\chi^2\left(\left[
\begin{array}{cc}
\cos\theta & -\sin\theta \\
\sin\theta & \cos\theta
\end{array}
\right]^T\left[u_1\ u_2\right]\right)}.
\end{equation}
The statistic $\chi_{\max}^2$ is used to detect an anomalous subgraph.

When the spectral norm is a reliable detection statistic, thresholding along the principal eigenvalue is often an effective method to identify the vertices that are exhibiting anomalous behavior. Working in multiple dimensions, while it enables the detection of smaller subgraphs, makes the process of identification more complicated. In this setting, we use a method based on $k$-means clustering to identify the subgraph vertices. Within the 2-dimensional space, we compute a small number of clusters, 
and declare the smallest cluster with at least a minimum number of vertices to be the signal subgraph.

\subsection{Eigenvector $L_1$ Norms}
\label{subsec:L1}
It is also desirable to detect signal subgraphs that do not stand out in the principal two components of the residuals matrix, and extending the algorithm of Section \ref{subsec:chi2} to an arbitrary number of dimensions may not be feasible. 
One method to detect such anomalies relies on the subgraphs being separable in the space of a single eigenvector. As mentioned previously, the entries in the eigenvectors of the background alone resemble numbers drawn from a Laplace distribution. Thus, if a subgraph were to stand out in a single eigenvector, that eigenvector will have a substantially smaller $L_1$ norm than for the background alone. The $L_1$ norm of a vector $x$, $\|x\|_1=\sum_{i}{|x_i|}$, is much smaller when it is concentrated on a small subset of entries, provided that it is unit-normalized in an $L_2$ sense. 
 For this reason, the $L_1$ norm serves as a proxy for sparsity in applications such as compressed sensing \cite{Don06}.

\begin{figure}
\begin{centering}
\includegraphics[width=1.72in]{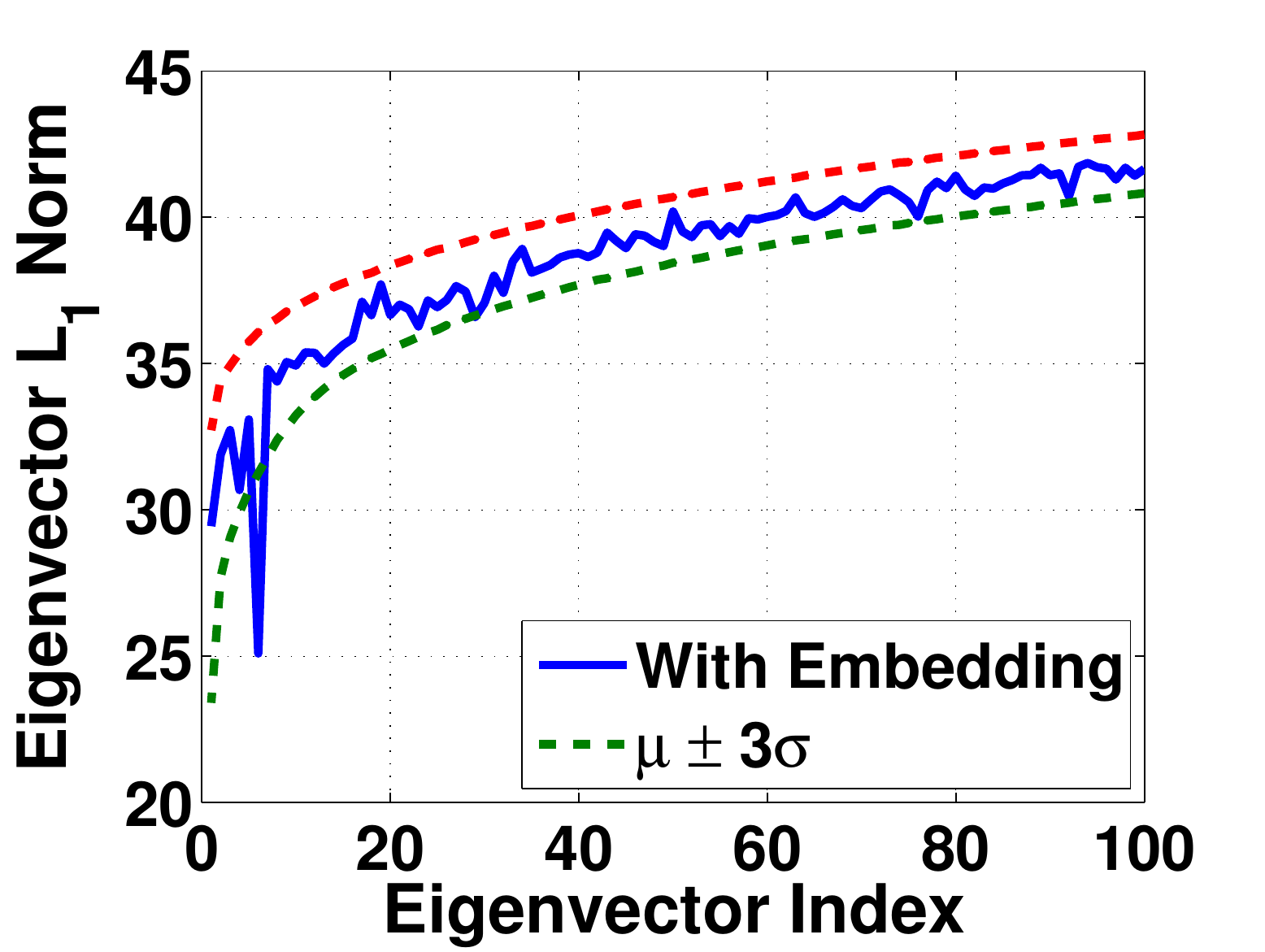}\ 
\includegraphics[width=1.72in]{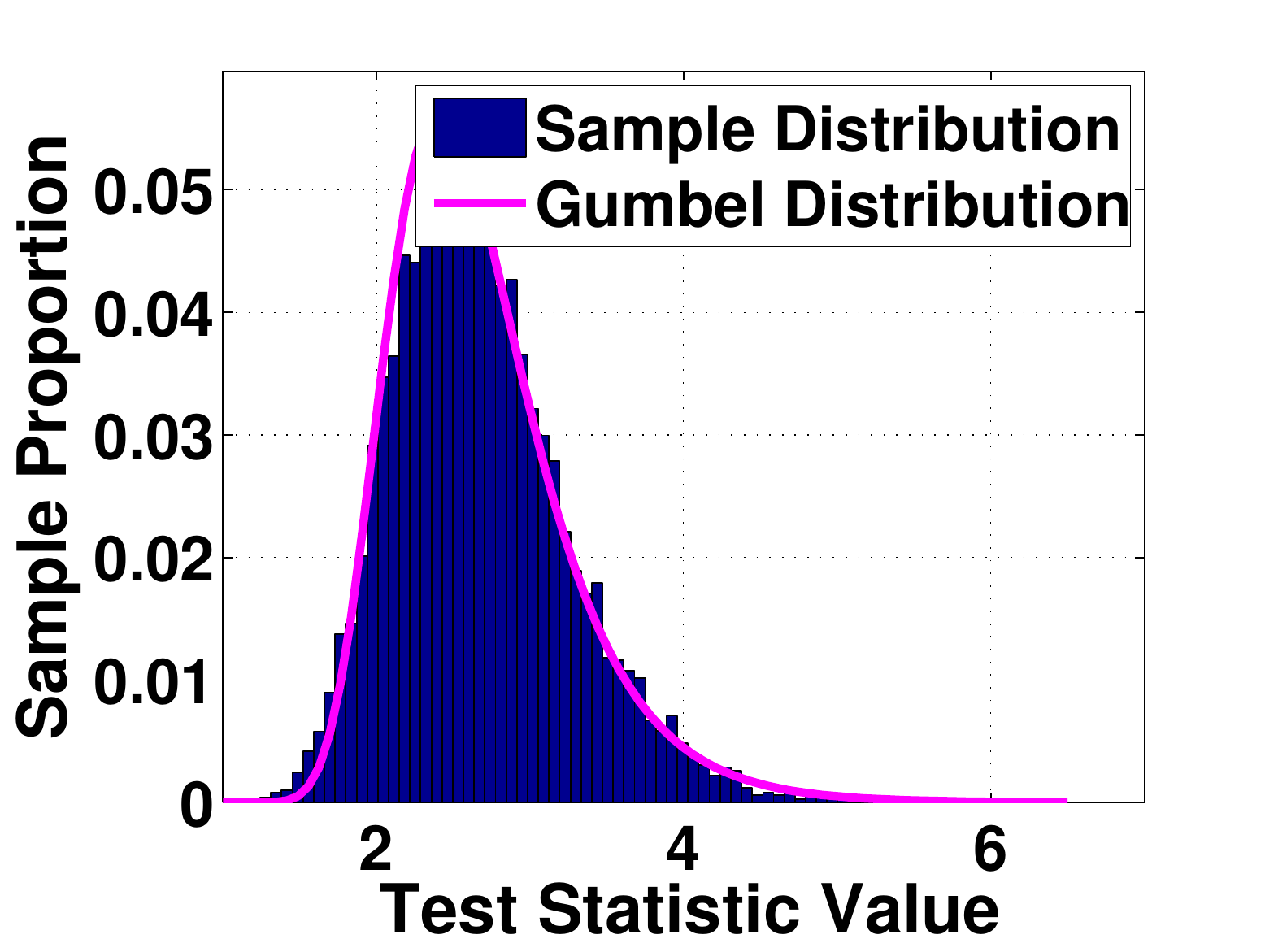}
\caption{An example of using eigenvector $L_1$ norms for subgraph detection. When a small, dense subgraph is embedded into a background with a skewed degree distribution, the $L_1$ norm of one of the eigenvectors of the residuals matrix becomes much smaller than usual, as shown on the left. Under the null hypothesis, the largest negative deviation from the mean will resemble a Gumbel distribution, plotted on the right.}
\label{fig:L1NormOutlier}
\end{centering}
\end{figure}

The following algorithm enables detection when an eigenvector is concentrated on the vertices of the subgraph. This will occur when, for example, a dense subgraph is embedded on relatively low degree vertices, as discussed in Appendix \ref{app:L1}. We compute the eigenvectors corresponding to the $m$ largest eigenvalues. By measuring cases with no embedding present, we obtain the mean $\mu_i$ and standard deviation $\sigma_i$ for the $L_1$ norm of the $i$th eigenvector. For each of the eigenvectors $u_i$, $1\leq i \leq m$, we subtract the mean and normalize by the standard deviation. The smallest (i.e., most negative) value is then used as a test statistic, since we are interested in cases where the norm is small. The test statistic is given by
\begin{equation}
-\min_{1\leq i\leq m}{\frac{\|u_i\|_1-\mu_i}{\sigma_i}}.\label{eqn:L1stat}
\end{equation}
An example demonstrating this method is provided in Fig. \ref{fig:L1NormOutlier}. The example uses a 4096-vertex graph with a skewed degree distribution (using the CL model described in Section \ref{subsubsection:chungLu}), with a 15-vertex subgraph with average degree 10.5 randomly embedded into the background. The analysis is run on the 100 eigenvectors associated with the largest positive eigenvalues. While the $L_1$ norms of most eigenvectors in the resulting matrix fall within 3 standard deviations of the mean for their index, the $L_1$ norm of the 6th eigenvector is over 10 standard deviations below the mean, which is extremely unlikely to occur under the null hypothesis. Under the null hypothesis, the test statistic (\ref{eqn:L1stat}) will resemble a Gumbel distribution (commonly used to model extreme values), as shown in the plot on the right. When an embedding occurs that creates a deviation as large as that in the lefthand plot, it will take on a value much larger than the maximum under normal circumstances.

The occurrence of tightly connected subgraphs highly aligned with eigenvectors was documented independently in \cite{Eigenspokes}, and  a similar anomaly detection method using eigenvector kurtosis in \cite{WWLZKurtosis}. Here, we use this phenomenon to find subgraphs whose internal connectivity is much larger than the expectation, given the background model. When an anomaly is detected according to (\ref{eqn:L1stat}), the corresponding eigenvector is thresholded to determine the subgraph vertices.

\subsection{Sparse Principal Component Analysis}
\label{subsec:SPCA}
While analysis of eigenvector $L_1$ norms enables the detection of some subgraphs that do not separate in the principal components of the residuals space, this technique has some shortcomings. In particular, as consecutive eigenvalues get closer together, the stability of the direction of the eigenvectors becomes unstable. Therefore, we cannot rely on the test statistic being sufficiently changed because an eigenvector points in the direction of the subgraph.

There is, however, a similar technique that enables the detection of small subgraphs with large residuals. Rather than first computing the eigenvectors of the residuals matrix and then finding an eigenvector with a small $L_1$ norm, we can find a vector that is \emph{nearly} an eigenvector whose $L_1$ norm is constrained. This is a technique known as sparse principal component analysis (sparse PCA) 
\cite{SDPforPCA}. This method has been used in the statistics literature to find high variance in the space of a limited number of variables. We utilize it here for a similar goal: to find large residuals in the space of a small number of vertices.

The problem is formulated as follows. The goal is to find a vector that is projected substantially onto itself by the residuals matrix, but with few nonzero components. Put formally, the objective is to solve
\begin{align}
&\xhat=\argmax_{\|x\|_2=1}{x^TBx}\\
&\textrm{subject to }\|x\|_0\leq N_S,\nonumber
\end{align}
where $\|\cdot\|_0$ denotes the $L_0$ quasi-norm (the number of nonzero components in a vector). This, however, is an integer programming problem, and is NP-hard. We therefore use a relaxation with an $L_1$ constraint, recast as a penalized optimization:
\begin{align}
&\xhat=\argmax_{\|x\|_2=1}{x^TBx-\lambda\|x\|_1}.
\end{align}
This problem is still not in an easily solvable form, due to the quadratic equality constraint. We use an additional relaxation, following the method of \cite{SDPforPCA}, to achieve a semidefinite program that can be solved using well-documented techniques:
\begin{align}
&\Xhat=\argmax_{X\in S_n}{\tr(BX)-\lambda\ones^T|X|\ones}\label{eqn:SDP}\\
&\textrm{subject to }\tr(X)=1,\nonumber
\end{align}
where $\tr(\cdot)$ denotes the matrix trace and $S_n$ is the set of positive semidefinite matrices in $\reals^{n\times n}$. The principal eigenvector of $\Xhat$, denoted $\xhat$, is then returned (and should be sparse, given the constraints). 
The subgraph detection statistic is $\|\xhat\|_1$. 
If no small subgraph has sufficiently large residuals, the vector should be relatively diffuse, and have a relatively large $L_1$ norm. For vertex identification, the sparse principal component is thresholded, and the vertices corresponding to the components of the vector greater than the threshold are declared to be part of the anomalous subgraph.

One drawback of this technique is its computational complexity. As mentioned in the introduction, one goal of this work is to develop techniques that scale to very large graphs. The algorithms described in Sections \ref{subsec:chi2} and \ref{subsec:L1} rely on a partial eigen\-decomposition. Using the Lanczos method for computing $m$ eigenvectors and eigenvalues of a matrix, and leveraging sparseness of the graphs, this requires a running time of $O((Mm+Nm^2+m^3)h)$, where $h$ is the number of restarts in the algorithm \cite{IRLanczos}. Thus, if the number of eigenvectors to compute is fixed, this algorithm scales linearly in the number of edges in its per-restart running time. Sparse PCA, as described in \cite{SDPforPCA}, has a running time that is $O(N^4\sqrt{\log{N}}/\epsilon)$, where $\epsilon$ controls accuracy of the solution. This implies that sparse PCA will not scale to extremely large datasets without additional optimization, which is a problem for future work. We present results using this technique to demonstrate the feasibility of detecting exceptionally small anomalies using the framework outlined in this paper.

\section{Simulation Results}
\label{sec:results}
\subsection{Noise Models}
\label{subsec:noise}
There are many models for random graphs, with varying degrees of complexity. In this section we outline 3 random models that will be used for background noise in our experiments.
\subsubsection{Erd\H{o}s--R\'{e}nyi (ER) Random Graphs}
\label{subsubsection:erdosRenyi}
The simplest random graph model was proposed by Erd\H{o}s and R\'{e}nyi in \cite{ER}. In this model, given a vertex set $V$ and a number $p\in(0,1)$, an edge occurs between any two vertices in $V$ with probability $p$. In matrix form, $p_{ij}=p$ for all $i$ and $j$. 
This model is subsumed by the model for a random graph with a given expected degree sequence assumed by equation (\ref{eqn:modMax}), where, in this case, all vertices have the same expected degree. 

\subsubsection{Chung--Lu (CL) Random Graphs}
\label{subsubsection:chungLu}
The ``given expected degree'' model has been studied extensively by Chung and Lu \cite{RandomGraphSpectra}. Similarly to the dynamic preferential attachment model of \cite{BAPA99}, in this model, the probability of two nodes sharing a connection increases with their popularity. Formally, each vertex $v_i$ is given an expected degree $d_i$, and the probability of vertices $v_i$ and $v_j$ sharing an edge is given by $p_{ij}=(d_id_j)/\sum_{\ell=1}^{|V|}{d_\ell}$, 
yielding a rank-1 probability matrix
\begin{equation}
P=\frac{1}{\sum_{i=1}^{|V|}{d_i}}dd^T.
\end{equation}
Using the observed degree as the expected degree---shown to be an approximately asymptotically unbiased estimator in \cite{ICASSP2012_Nick}---the standard formulation of the modularity matrix (\ref{eqn:standardModMat}) perfectly fits this model for background behavior.

\subsubsection{R-MAT Stochastic Kronecker Graphs}
To include a slightly more complicated model, we also consider the Recursive Matrix (R-MAT) stochastic Kronecker graph \cite{RMAT}. In this model, a base probability matrix
\begin{equation}
P_b=\left[\begin{array}{cc}
a & b\\
c & d\end{array}\right]
\end{equation}
is given, where $a$, $b$, $c$ and $d$ are nonnegative values that sum to 1, and edge probabilities are defined by the $n$-fold Kronecker product of $P_b$, denoted $\widehat{P}=\{\hat{p}_{ij}\}=\bigotimes_{i=1}^{n}{P_b}$. This results in matrices with $2^n$ vertices. The graph is generated by an iterative method where 1 edge is added at each iteration with probabilities defined by $\widehat{P}$. If the total number of iterations is $t$, the edge probabilities are given by
\begin{equation}
\label{eqn:rmatProb}
p_{ij}=1-(1-\hat{p}_{ij})^t.
\end{equation}
If the base probability matrix has rank 1, this generator will produce graphs with a similar structure to the CL model. As shown in \cite{RMAT}, however, this model creates graphs with mild community structure, thereby presenting a more challenging noisy background for our subgraph detection framework.

These three models represent varying degrees of complexity for the detection framework. The ER model is overspecified by the given expected degree model used in the modularity matrix, the CL model matches the formula exactly, and the R-MAT model is mismatched due to its mild community structure. In the simulations in Section \ref{subsec:sims}, the R-MAT graphs are generated using a base probability matrix with $a=0.5$, $b=c=0.125$, and $d=0.25$, 
and the algorithm is run for $12N$ iterations, resulting in an average degree of approximately 12. The graph is unweighted, and undirected via the ``clip-and-flip'' procedure as in \cite{RMAT}, i.e., the edges below the main diagonal in the adjacency matrix are removed, and those above the main diagonal are made undirected. For CL backgrounds, the expected degree sequence is defined by the edge probabilities of the R-MAT background, i.e., $d_i=\sum_{j=1}^{|V|}{p_{ij}}$ where $p_{ij}$ is defined as in (\ref{eqn:rmatProb}). The ER backgrounds use an edge probability that yields an average degree the same as the more complicated models.

\begin{figure}
\begin{centering}
\hspace*{-0.05in}\includegraphics[width=1.2in]{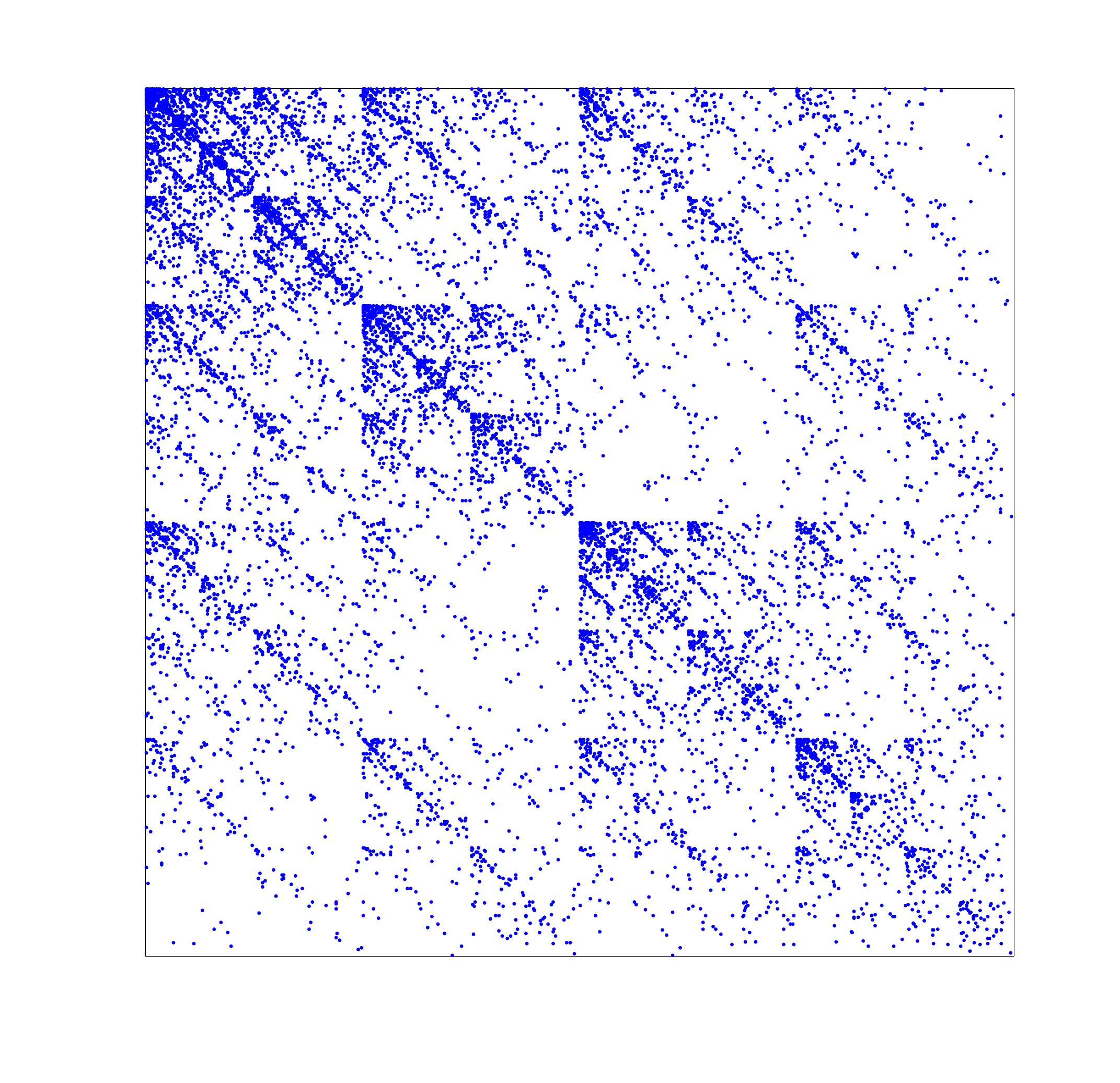}
\includegraphics[width=1.2in]{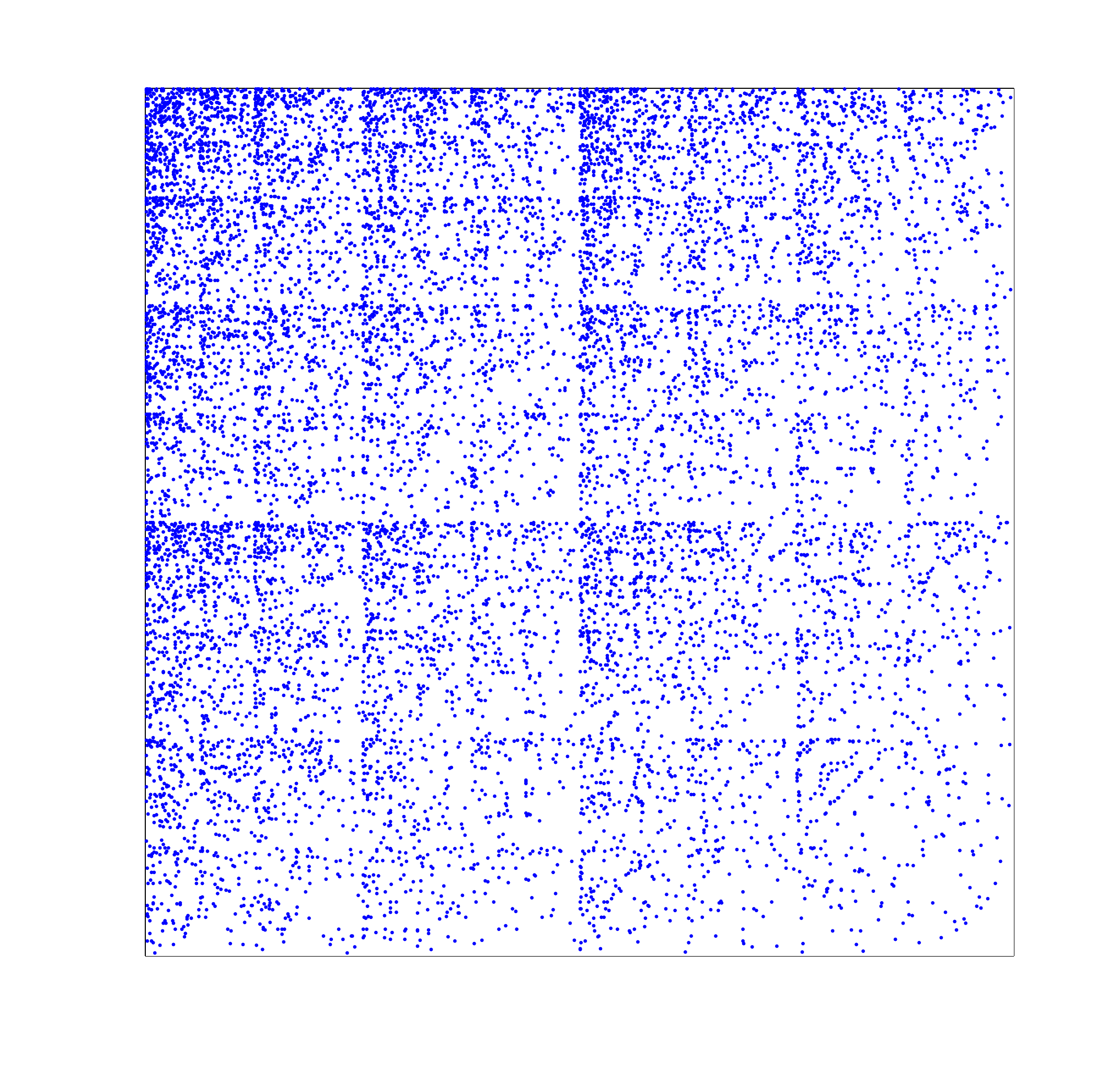}
\includegraphics[width=1.2in]{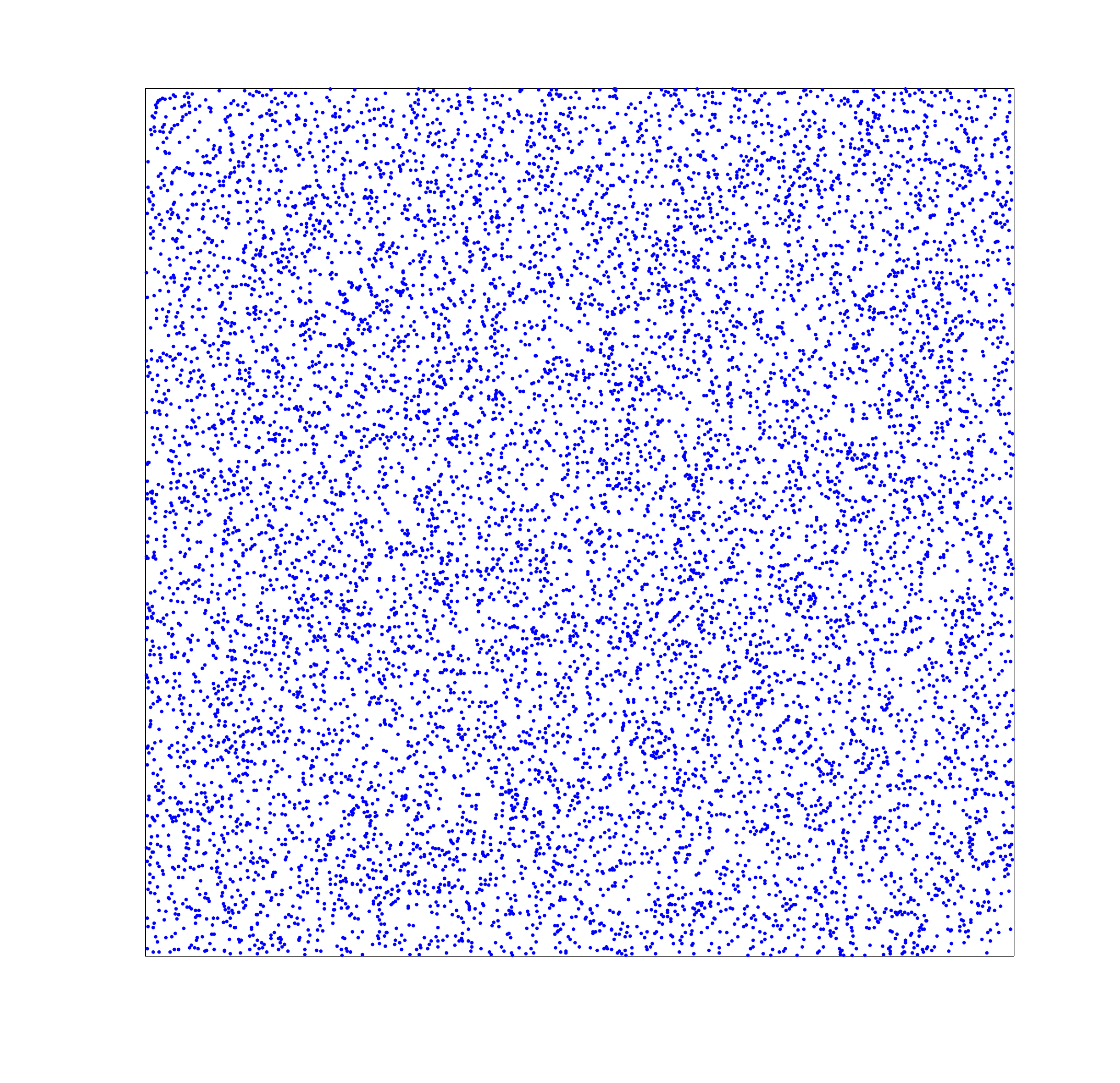}
\caption{Sparsity patterns for background graphs: an R-MAT graph (left), a Chung--Lu graph (center) and an Erd\H{o}s--R\'{e}nyi graph (right).}
\label{fig:bgAdj}
\end{centering}
\end{figure}

Example sparsity patterns of the adjacency matrices, each with 1024 vertices, are shown in Fig. \ref{fig:bgAdj}. Note the moderate community structure in the R-MAT graph. While the CL graph has vertices of varying degree, it does not have the same structure of the R-MAT. One particularly visible difference is the lack of connections between low-degree vertices and high-degree vertices in the R-MAT graph, seen in the upper-right and lower-left corners of the matrix. Both of these graphs contain more variation that the ER graph, where the uniform randomness can be seen in its sparsity pattern.

\subsection{Signal Subgraph}
Two random graph models are used for the anomalous signal subgraph. In one case, an ER graph with probability parameter $p_S$ is generated and combined with randomly selected vertices from the background. Here, the expected adjacency matrix is an $N_S\times N_S$ matrix where every entry is $p_S$, and thus has spectral norm $p_SN_S$. The second subgraph we consider is a random bipartite graph, where the vertex set is split into two subsets and no edge can occur between vertices in the same subset. Letting $N_1$ and $N_2$ be the numbers of vertices in each subset, there are $N_1N_2$ possible edges between the two vertex subsets, and, as in the ER subgraph case, each of these possible edges is generated with equal probability $p_S$. For the bipartite subgraph, the expected adjacency matrix has the form
\begin{equation}
\expect\left[A_S\right]=\left[\begin{array}{cc}
{\bf 0}_{N_1\times N_1} & p_S{\bf 1}_{N_1\times N_2}\\
p_S{\bf 1}_{N_2\times N_1} & {\bf 0}_{N_2\times N_2}
\end{array}\right],
\end{equation}
which has spectral norm $p_S\sqrt{N_1N_2}$. This subgraph provides us with a signal where the average degree does not equal the spectral norm (unless $N_1=N_2$), demonstrating that the spectral norm is a more appropriate power metric.

\subsection{Monte Carlo Simulations}
\label{subsec:sims}
\begin{figure*}
\begin{centering}
\includegraphics[width=2.44in]{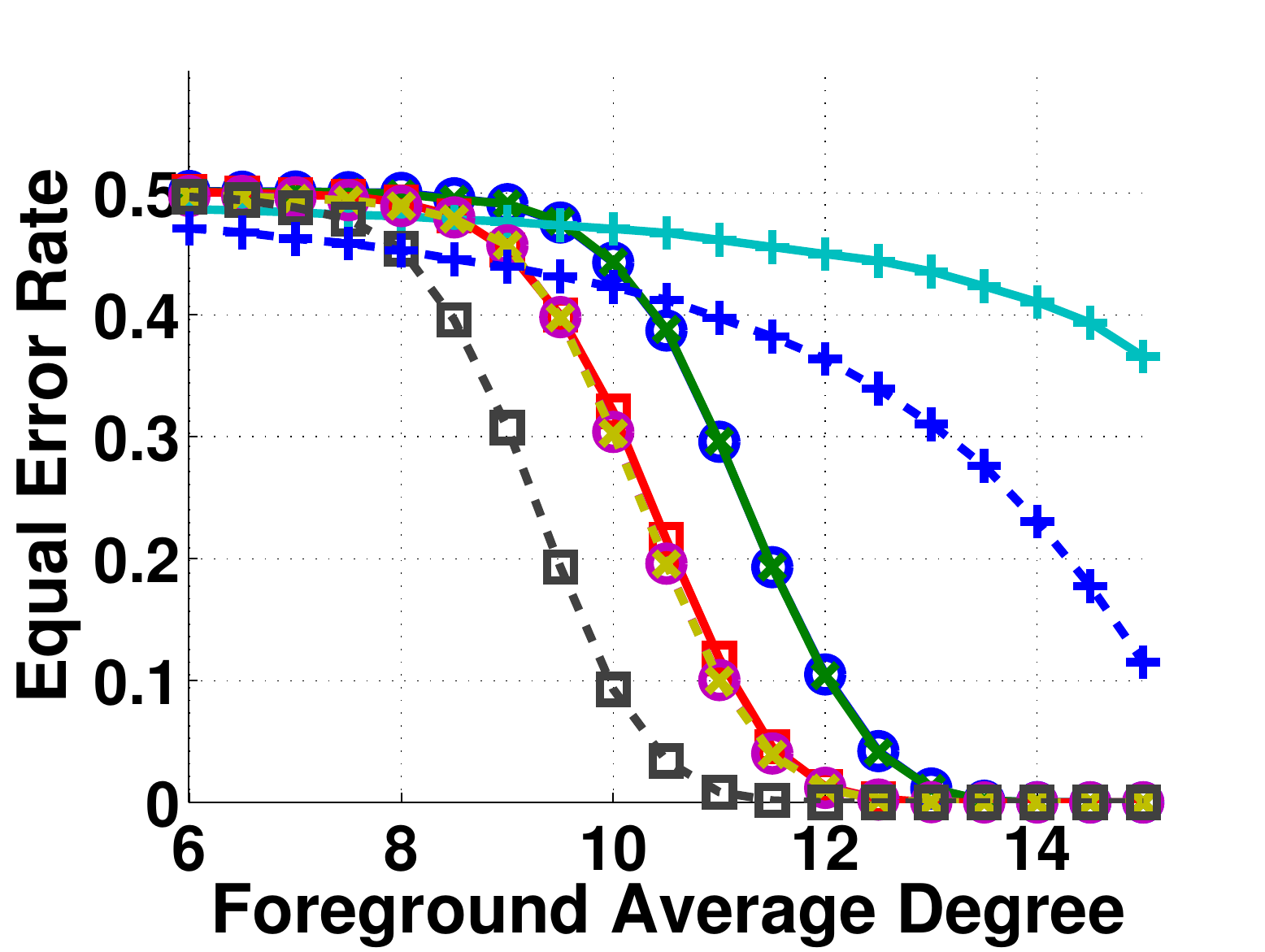}%
\includegraphics[width=2.44in]{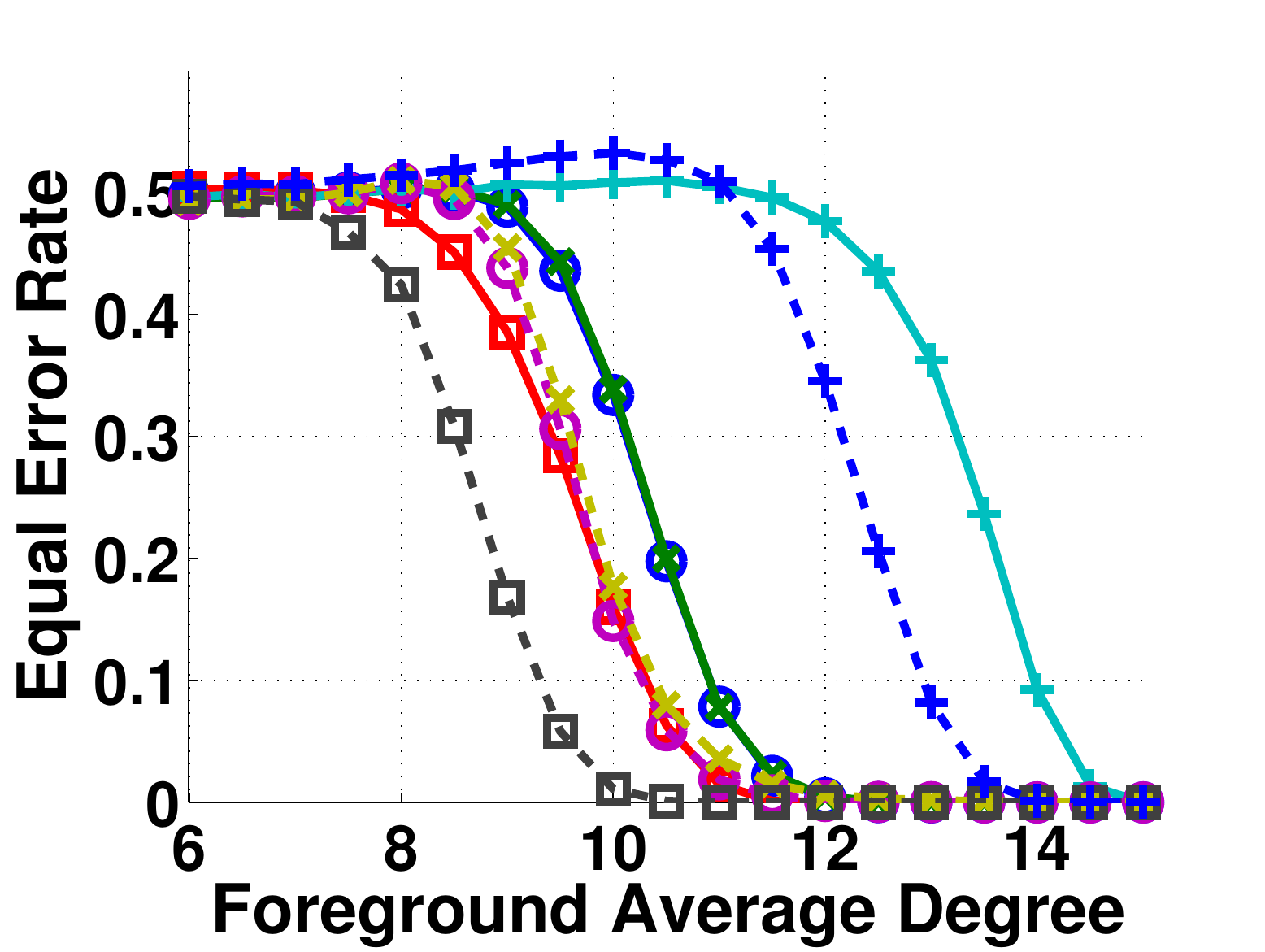}%
\includegraphics[width=2.44in]{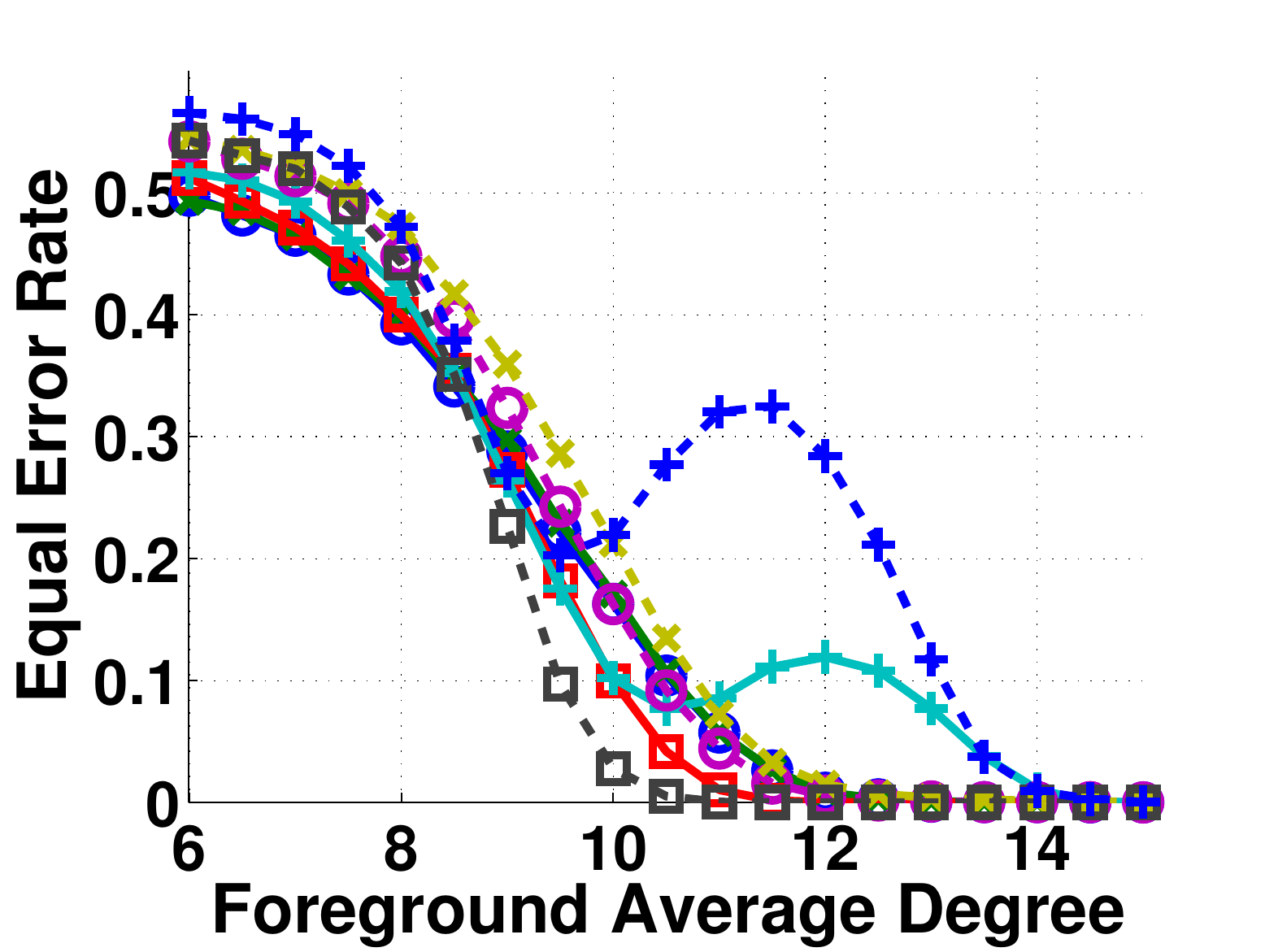}\\%
\includegraphics[width=2.44in]{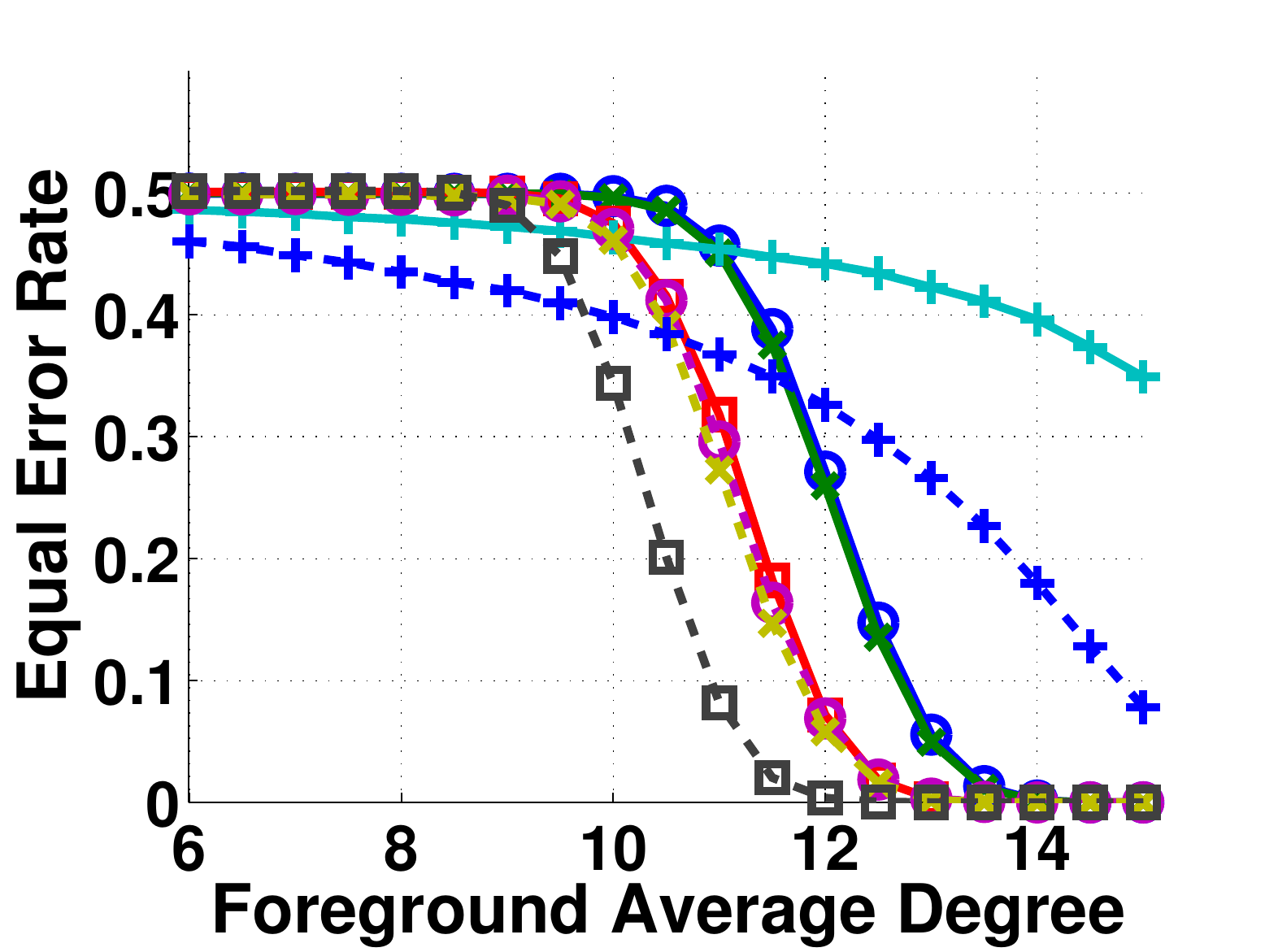}%
\includegraphics[width=2.44in]{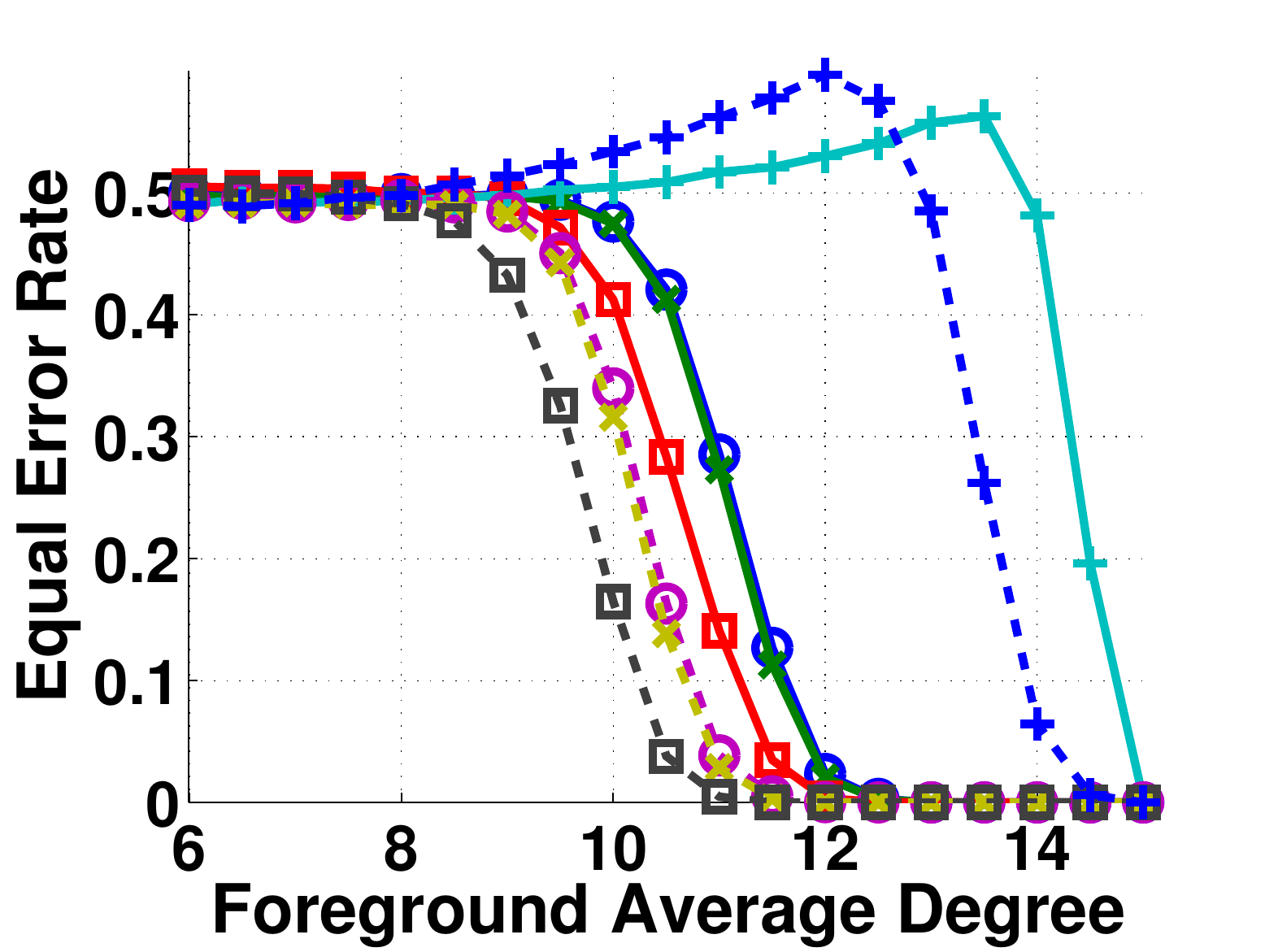}%
\includegraphics[width=2.44in]{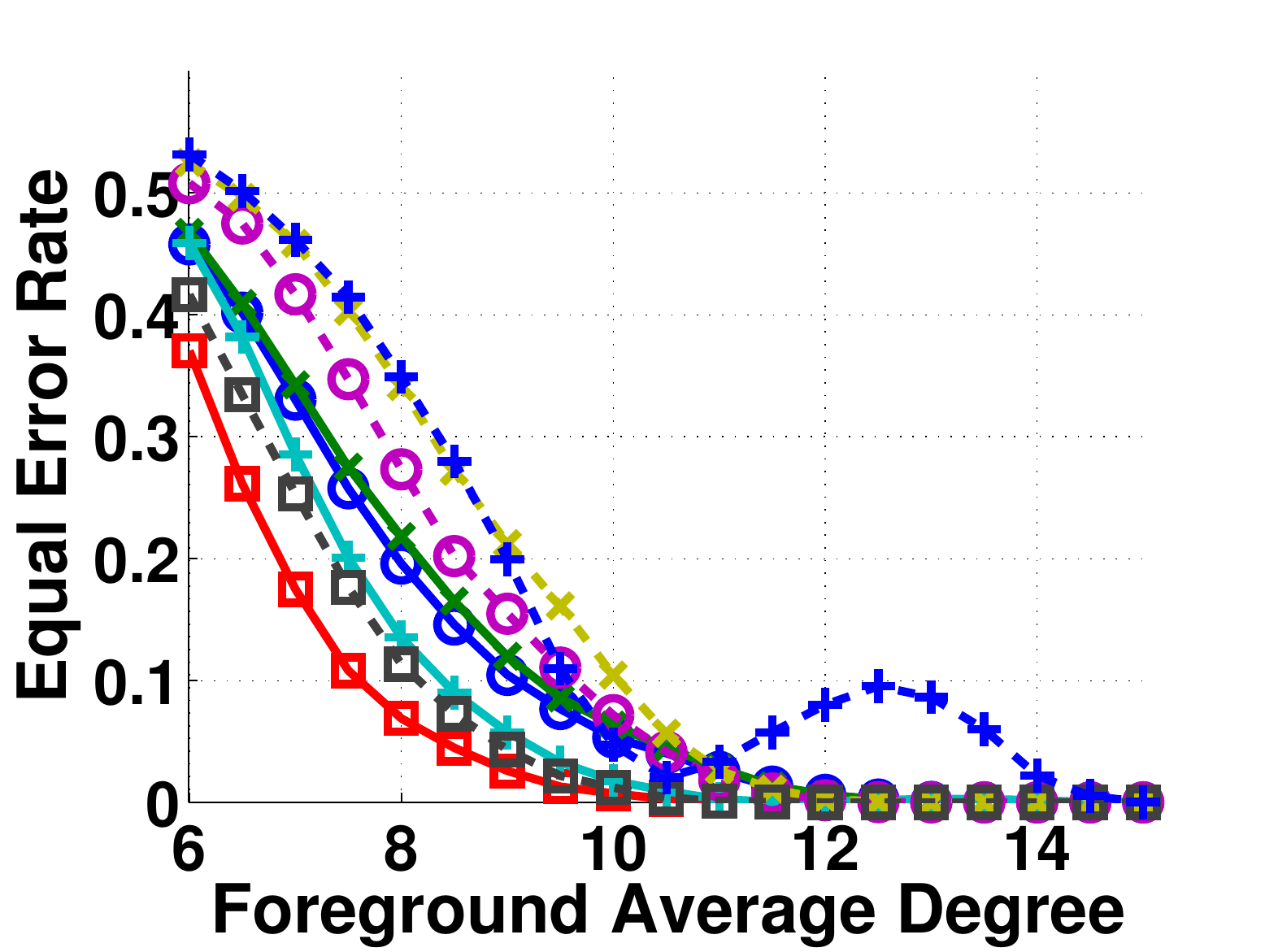}\\%
\includegraphics[width=2.44in]{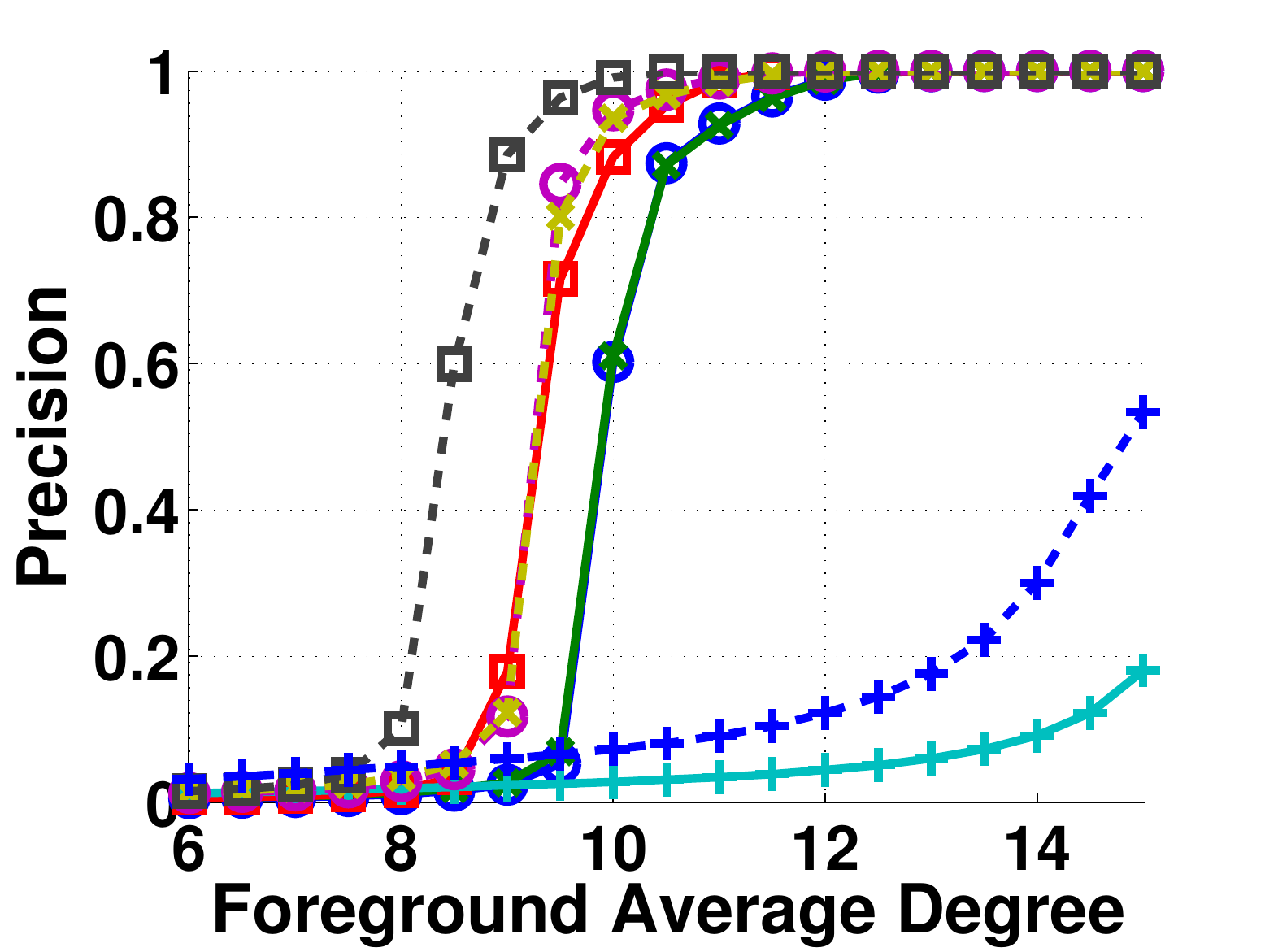}%
\includegraphics[width=2.44in]{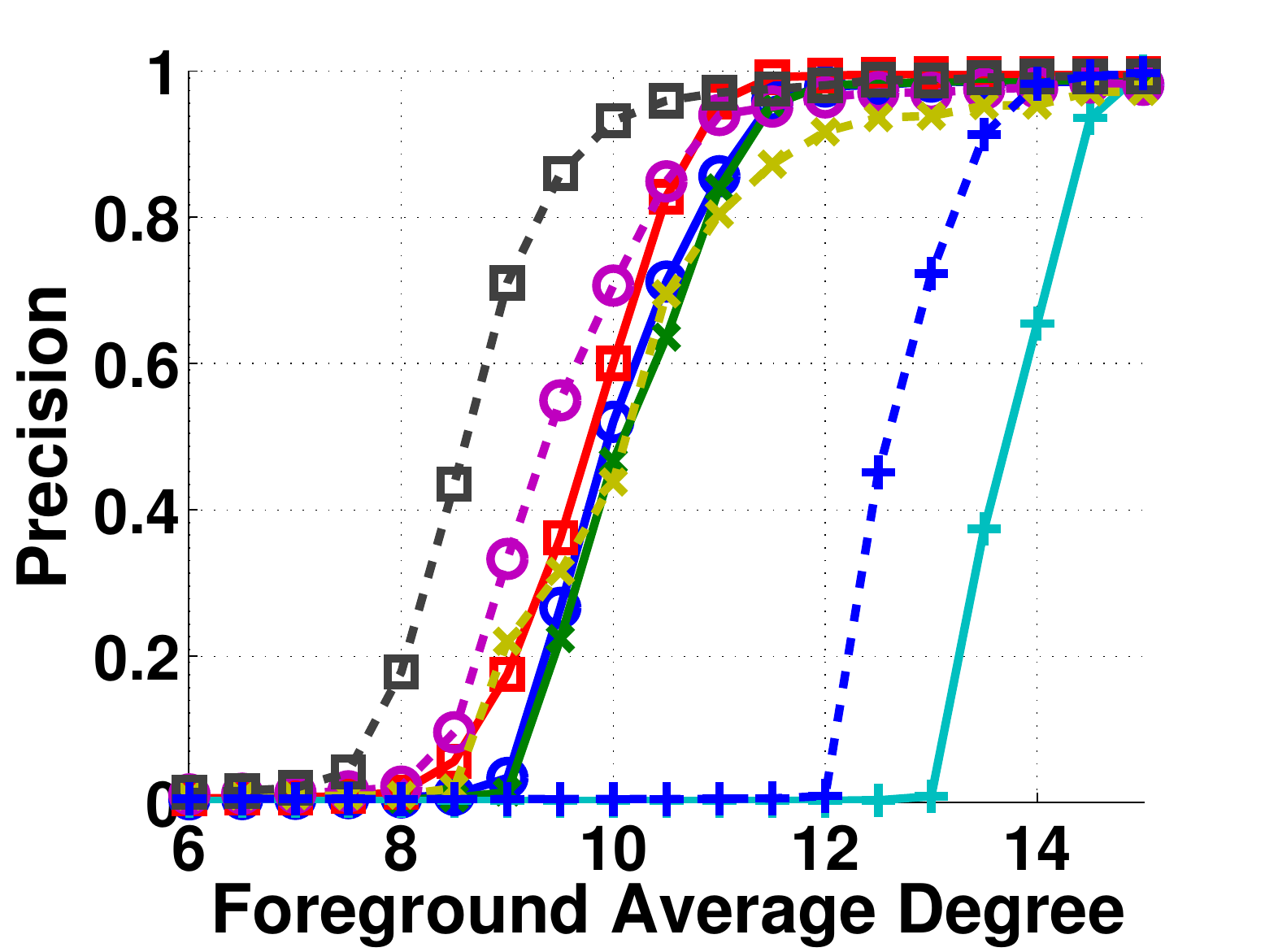}%
\includegraphics[width=2.44in]{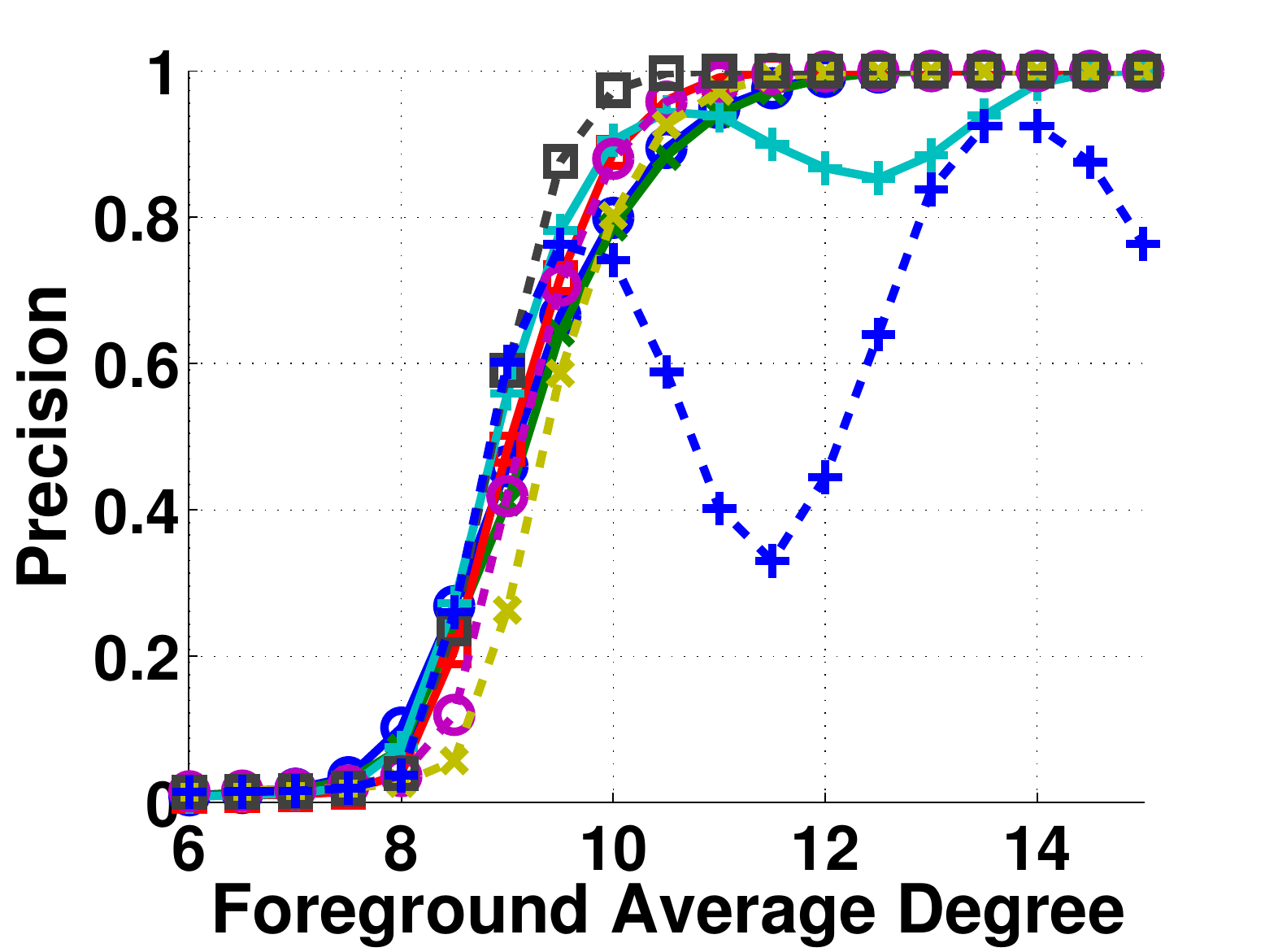}\\%
\includegraphics[width=2.44in]{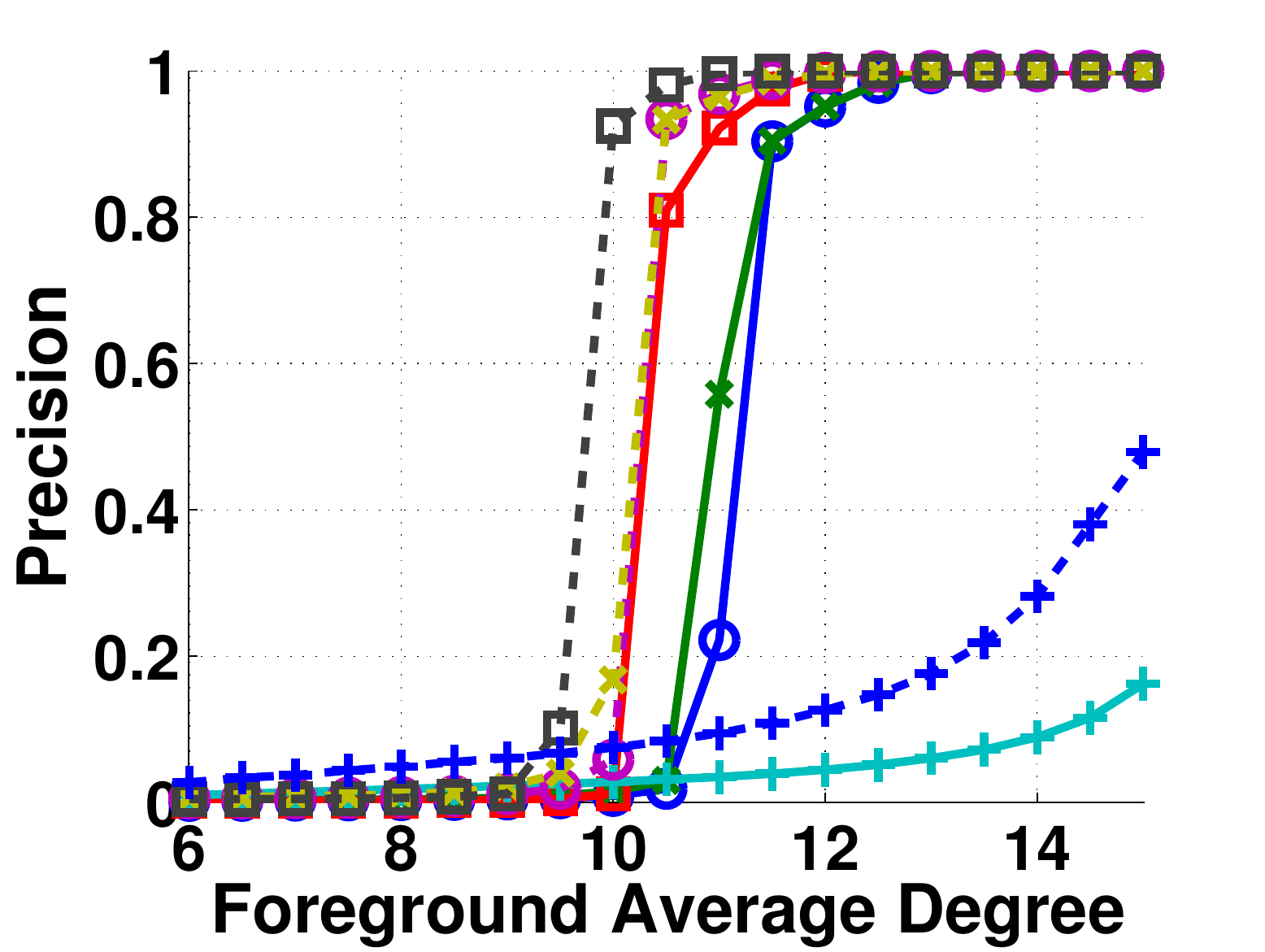}%
\includegraphics[width=2.44in]{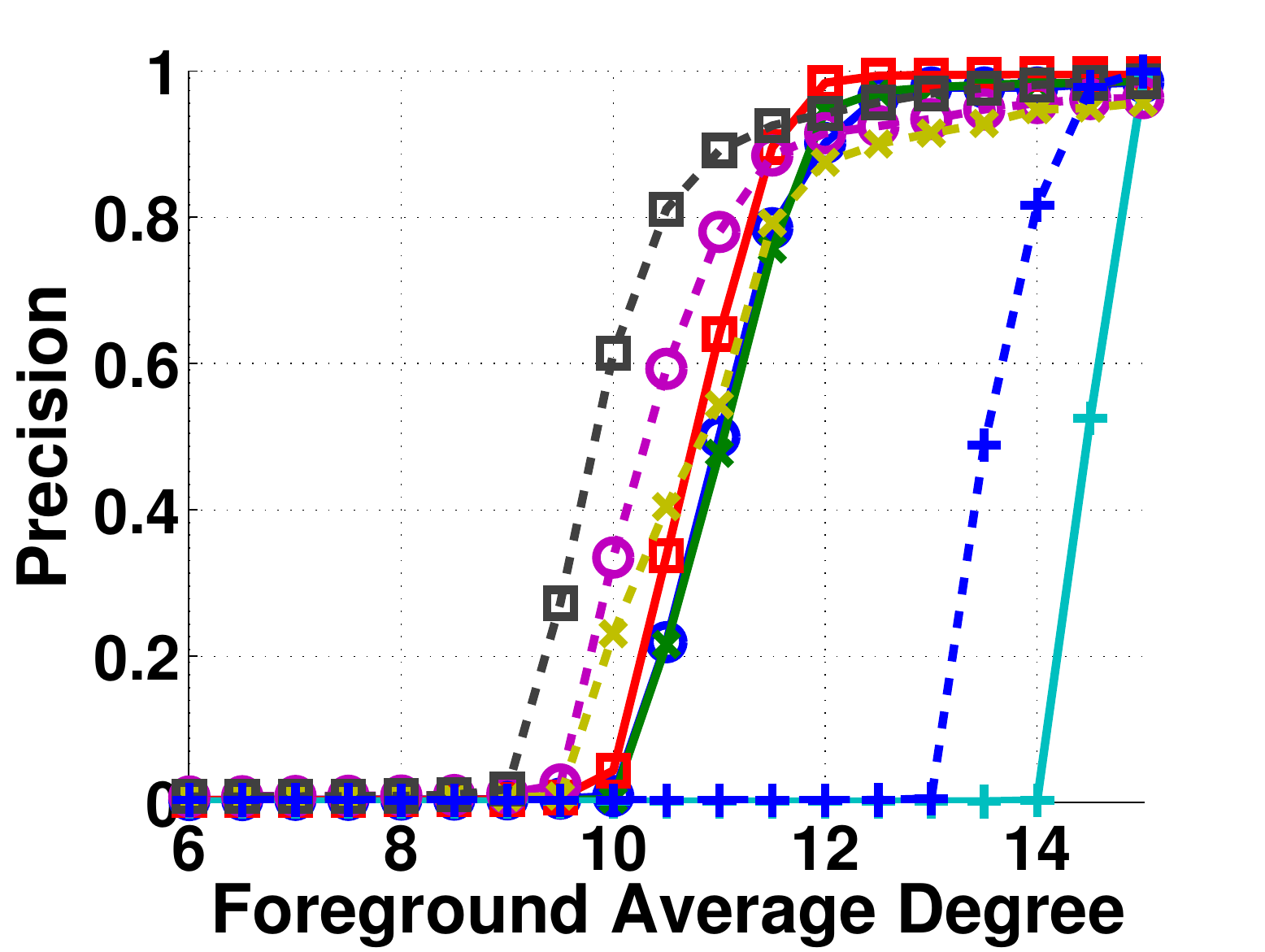}%
\includegraphics[width=2.44in]{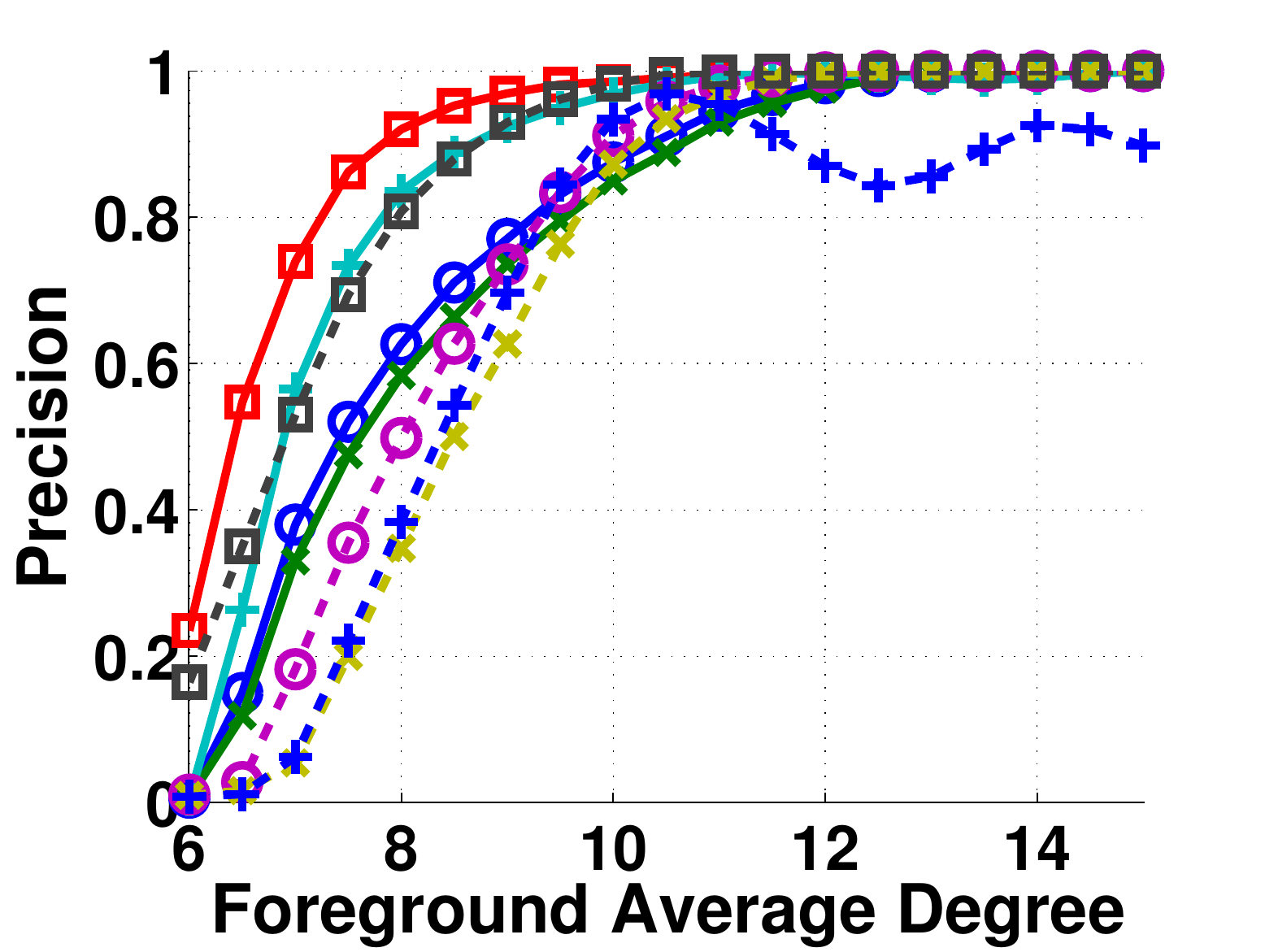}%
\caption{A summary of detection and identification performance. The equal error rate (EER) for each background and foreground is shown as the average degree increases from 6 to 15. Results are shown for cluster subgraphs (solid line) and bipartite subgraphs (dashed line), for R-MAT graphs with the true expected value ($\square$), R-MAT graphs with an estimated rank-1 expected value ($+$), CL graphs with given expected degrees ($\circ$) and CL graphs using observed degrees ($\times$). Performance improves as the test statistic goes from the spectral norm (left column), to the chi-squared statistic (center column) to the largest deviation in $L_1$ norm (right column). Detection performance with the $L_1$ norm-based statistic improves when the subgraph is embedded on low-degree vertices (second row), rather than choosing the vertices uniformly at random (first row). The same performance trends typically hold for the vertex identification algorithms (uniform random embedding in third row, degree-biased embedding in fourth row), shown here in terms of precision at a 35\% recall rate. The non-monotone behavior using $L_1$ norms is caused by a cluster of larger eigenvalues in the R-MAT background, which, as discussed in Appendix \ref{app:L1},  makes detection more difficult with this method.}
\label{fig:EER}
\end{centering}
\end{figure*}

The results in this section detail the outcomes of several 10,000-trial Monte Carlo simulations. In each simulation, a background graph is generated, and may or may not have a signal subgraph embedded on a subset of its vertices. The subgraph may be a 15-vertex cluster or a bipartite graph with $N_1=12$ and $N_2=25$. Test statistics outlined in Section \ref{sec:algorithms} are computed on the resulting graph, creating several empirical distributions that can be used to discriminate between $\calH_0$ and $\calH_1$. Residuals matrices are formed using either the exact expected value\footnote{Due to time and memory constraints, a rank-100 approximation for the R-MAT expected value was used instead of the true probability matrix.}, or a rank-1 approximation based on the observed degrees, as in (\ref{eqn:modMax}). 
The expected degree sequence from the R-MAT model is used for CL backgrounds, and ER backgrounds use the same average degree. For R-MAT and CL backgrounds, we consider cases where the foreground vertices are selected uniformly at random from all background vertices, and cases where they are randomly selected from the set of vertices with expected degree less than 5.

For 4096-vertex graphs, ER graphs always achieved near-perfect detection performance. Identification and detection performance for CL and R-MAT backgrounds are summarized in Fig.~\ref{fig:EER}. A few phenomena in the results confirm our intuition. 
First, note that CL backgrounds have extremely similar performance, whether the expected value term is given or estimated. This is because the observed degree is a good estimate for expected degree, and the small embedding has a minimal effect on the expected value term, as shown in Appendix \ref{app:deltaK}. (The small but noticeable difference when using a bipartite foreground emphasizes the impact of the number of subgraph vertices.) The R-MAT backgrounds have much more substantial performance differences, due to the model mismatch. In fact, when the true expected value is given, performance is better than with the CL background. This is likely due to the lower variance in the noise, caused by smaller connection probabilities among low-degree vertices. 
Detection performance improves going from the spectral norm statistic to the chi-squared statistic, and improves further when analyzing the eigenvector $L_1$ norms. Also, when the subgraph is embedded only on vertices with expected degree at least 5, performance significantly increases for $L_1$ norm analysis, while it degrades for the other statistics (since it is likely to be more orthogonal to the principal eigenvectors). 
Note also that, for the spectral norm and chi-squared statistics, the bipartite embedding is more detectable than the cluster with the same average degree, since the bipartite foreground has a higher spectral norm. This does not hold for the $L_1$ norm statistic, since the cluster embedding, while less powerful, is concentrated on a smaller subset of vertices, making it more detectable using this statistic. 

\begin{figure}
\begin{centering}
\includegraphics[width=3.25in]{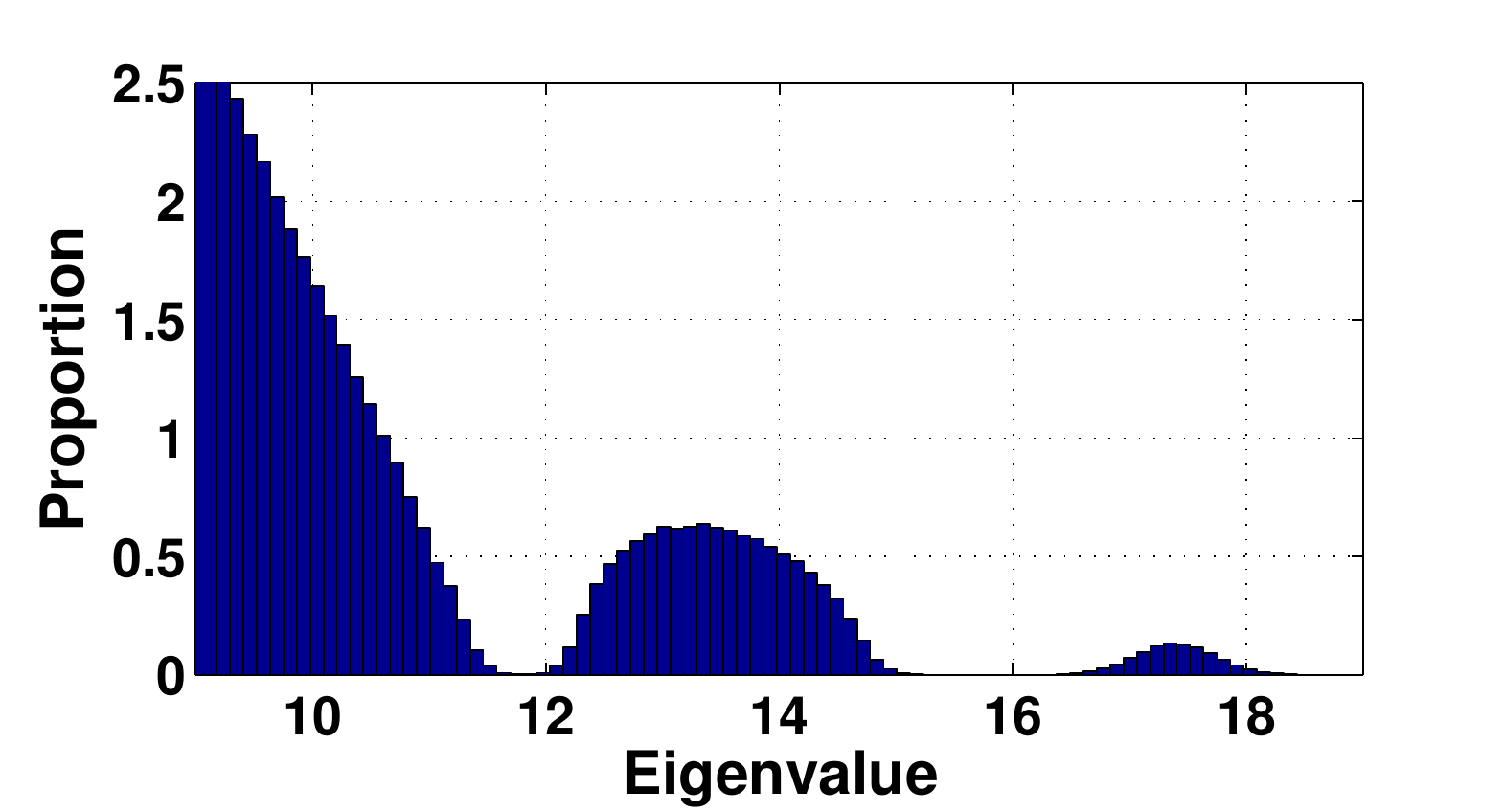}
\caption{Histogram of eigenvalues from an R-MAT matrix using an estimated rank-1 expected value. The two clusters of larger eigenvalues are responsible for the non-monotonic behavior in the $L_1$-norm statistic in shown in Fig. \ref{fig:EER}.}
\label{fig:RMATEVs}
\end{centering}
\end{figure}

One interesting aspect of the $L_1$ norm technique is its non-monotonic behavior when using the estimated rank-1 expected value. In both detection and identification, performance improves as the subgraphs increase in size up to a certain point, after which performance degrades and then improves again. This is due to clustering of eigenvalues caused by the model mismatch, as shown in Fig. \ref{fig:RMATEVs}. The figure presents a histogram of eigenvalues for the R-MAT graph minus the estimated rank-1 expected value matrix, $\|k\|_1^{-1}kk^T$. (The vertical axis is the average number of eigenvalues that fell into a given bin over the 10,000 Monte Carlo trials.) Most of the eigenvalues are below 12, while there is always 1 that is over 16 and 11 in the cluster that spans approximately 12 to 15. Since, as discussed in Appendix \ref{app:L1}, having eigenvalues that are close together hinders performance with this method, performance improves when the subgraph can be localized in an eigenvector as its eigenvalue approaches 12, but it will be more difficult around 13. Using the true expected value instead of the rank-1 approximation does not yield this behavior, since there is no model mismatch. The mismatch between R-MAT and the rank-1 expected value also causes the slight degradation in performance using the chi-squared statistic before it rapidly improves. This may be because the embedded subgraph actually improves the symmetry of the projection by balancing out the mismatch, before finally overpowering it.

The identification results on the bottom half of the figure follow similar trends, with one notable exception. Performance is shown in terms of precision at a 35\% recall rate (precision is emphasized since the foreground vertex set is much smaller than the background). While the $k$-means-based identification method (center column, using 3 clusters and a subgraph threshold of 5 vertices) typically improves performance over thresholding of the principal eigenvector (first column) for cases where precision is relatively low, it actually hinders performance in cases where precision is high. This shows that a subgraph that separates well along the first eigenvector will not necessarily be equally detectable via $k$-means, possibly due to spreading in the second dimension.

\begin{figure}
\begin{centering}
\includegraphics[width=1.70in]{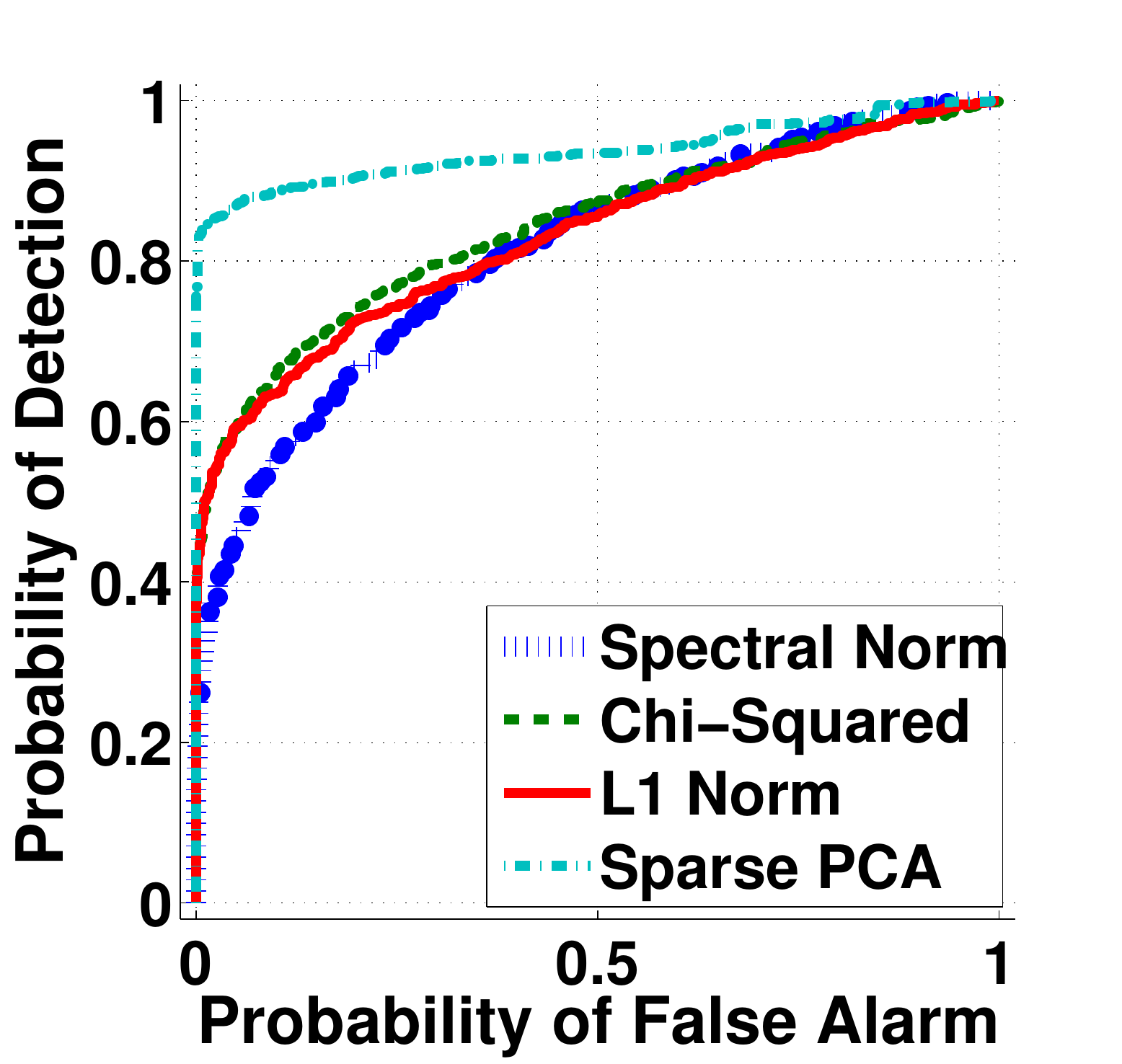}
\includegraphics[width=1.70in]{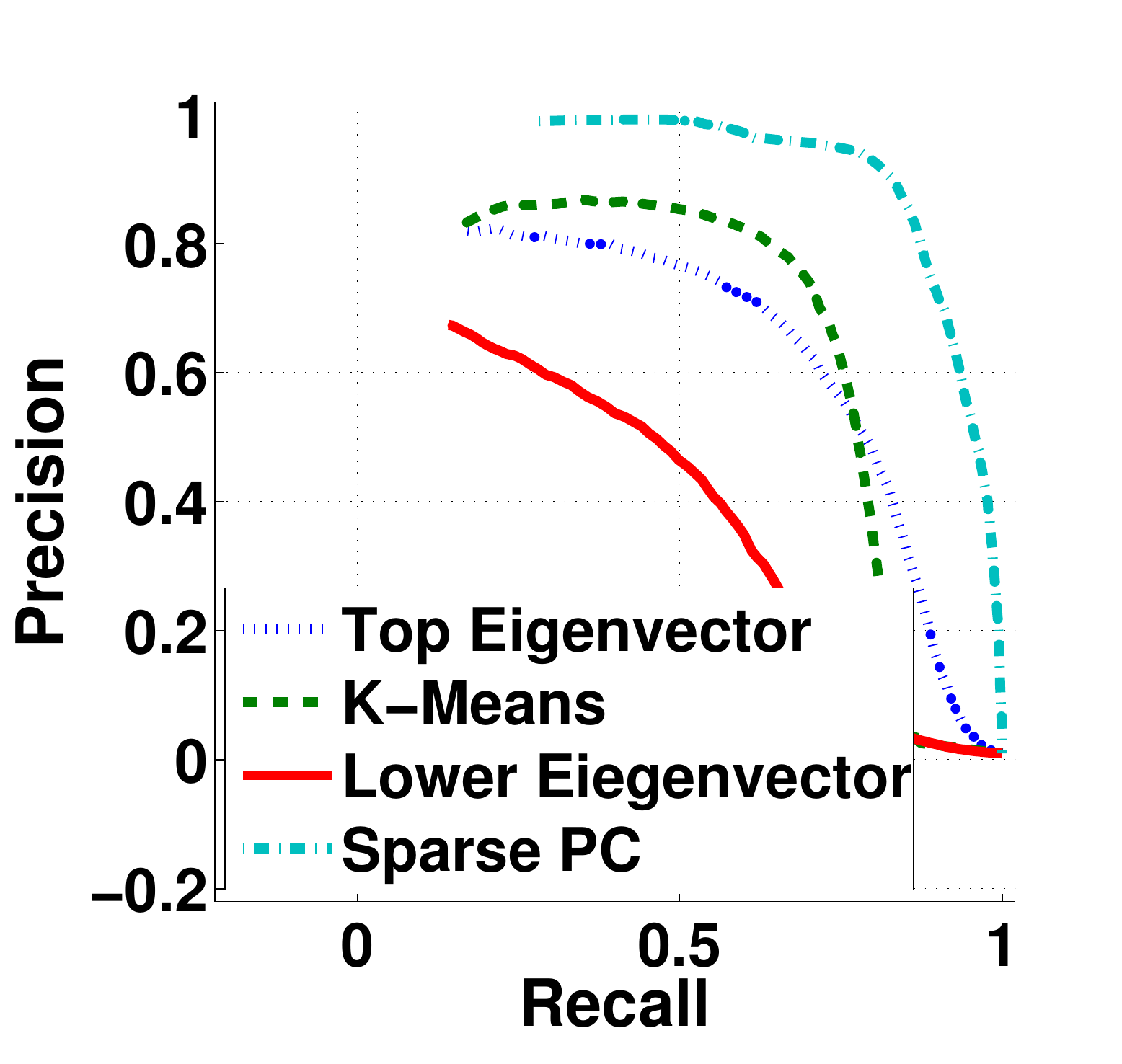}\\
\includegraphics[width=1.70in]{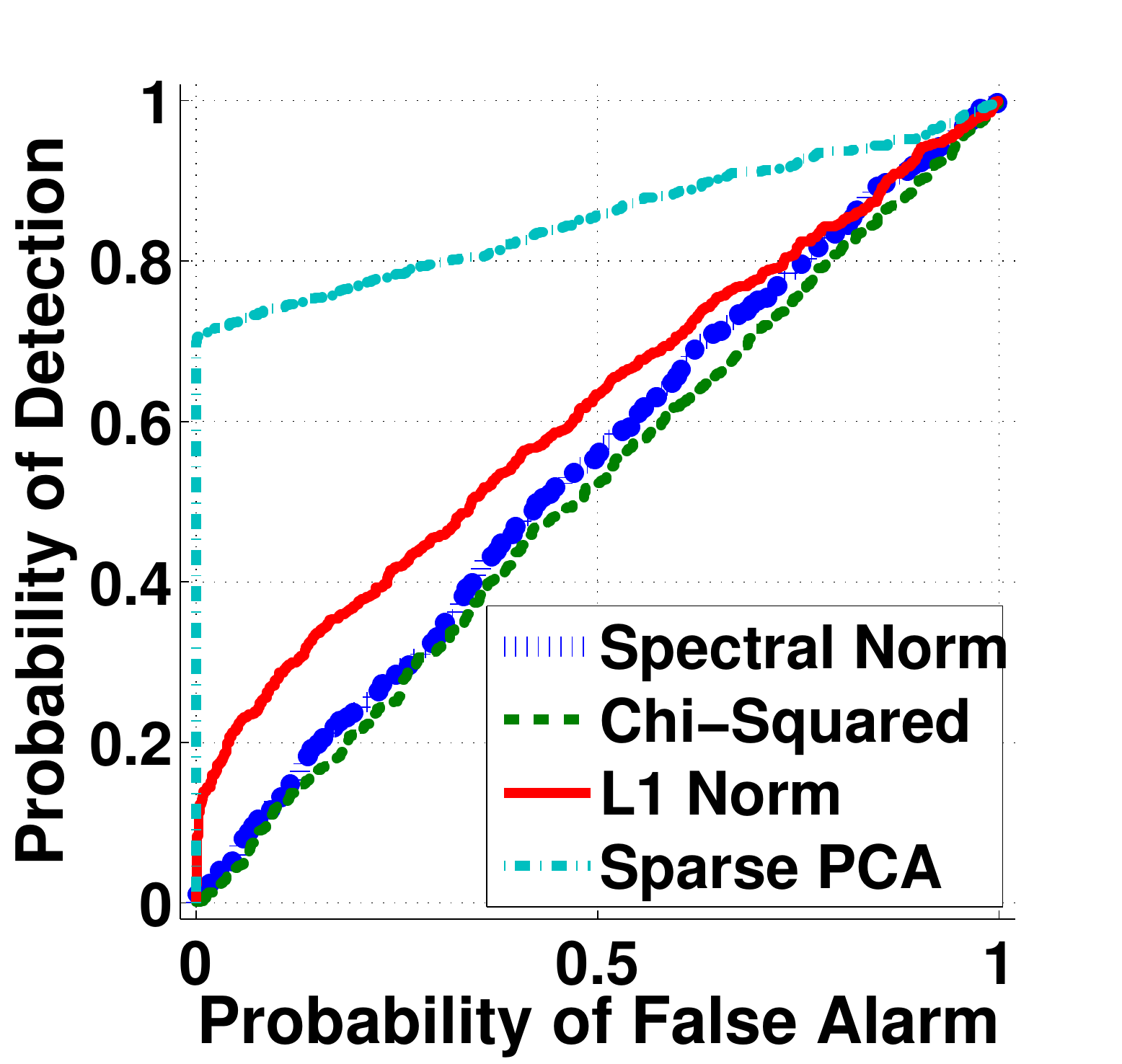}
\includegraphics[width=1.70in]{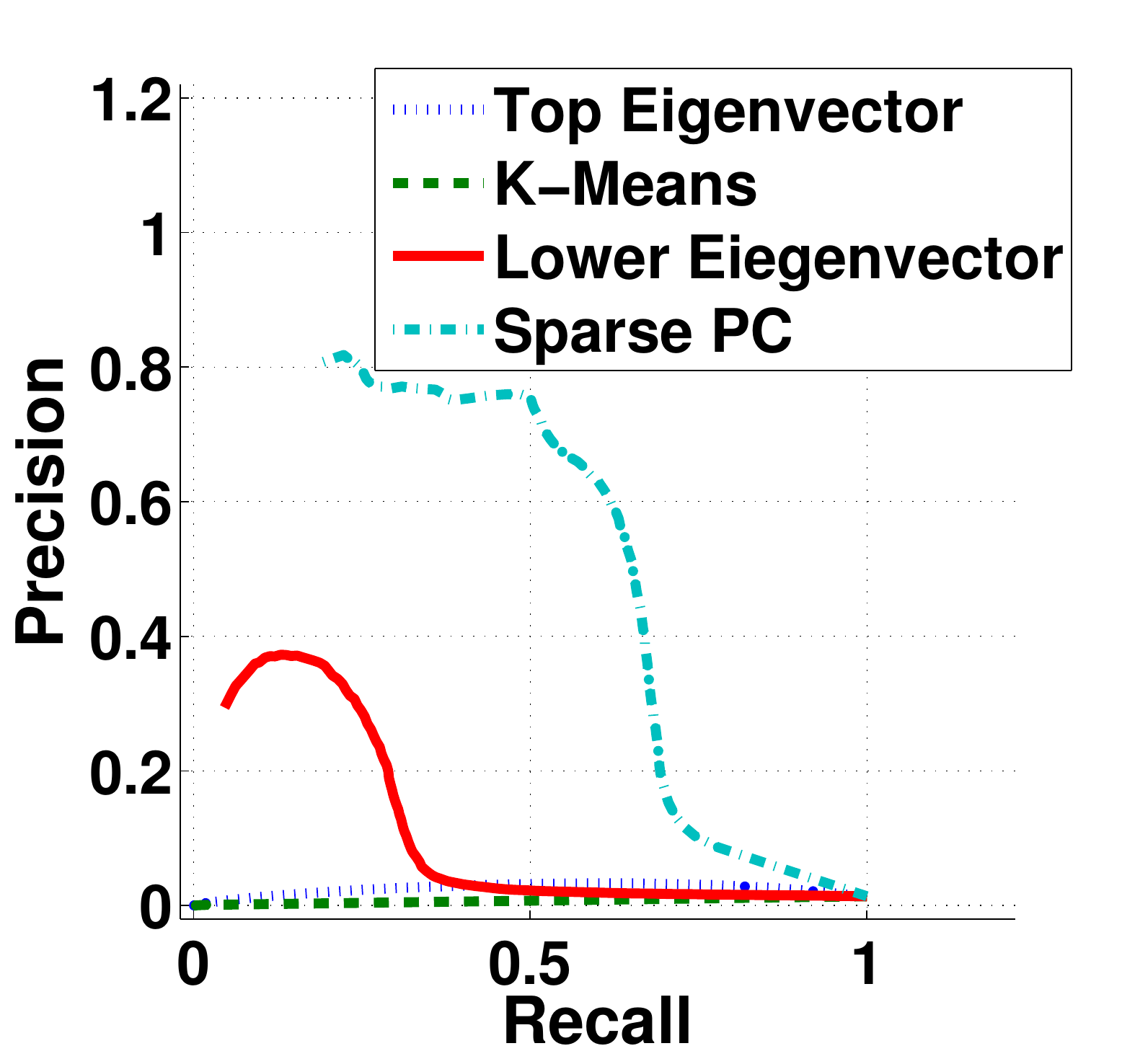}
\caption{Detection and identification results using sparse PCA. In both an Erd\H{o}s--R\'{e}nyi background (top row) and an R-MAT background (bottom row), sparse PCA significantly outperforms the other algorithms. Similar performance gaps are seen in detection performance (left column) and identification (right column).}
\label{fig:SPCA}
\end{centering}
\end{figure}

Since sparse PCA has a much greater computational burden, we carried out a more limited set of experiments on smaller graphs. In each trial, a 512-vertex background graph is generated according to either an R-MAT or ER model. The R-MAT graphs use the same probability matrix as in the previous experiment, and the ER graphs have equal expected volume. In each case, we use an estimated rank-1 expected value, and use the DSPCA software package \cite{DSPCA} to solve (\ref{eqn:SDP}). Detection and identification performance are shown in Fig. \ref{fig:SPCA}. These results demonstrate the detection of a 7-vertex, 80\% dense subgraph in the R-MAT background or a 5-vertex, 85\% dense subgraph in an ER background. 
Sparse PCA yields markedly superior performance to the three methods used in Fig. \ref{fig:EER}. By using this more costly technique, much smaller, subtler anomalies can be detected, using the same principles as the less expensive algorithms.

\section{Results on Application Data}
\label{sec:applications}
Two network datasets were downloaded from the Stanford Network Analysis Project (SNAP) large graph dataset collection 
(available at http://snap.stanford.edu/data). 
One dataset consists of product co-purchase records on amazon.com, where each of the 548,552 vertices represents a product, and a directed edge from vertex $i$ to vertex $j$ denotes that when product $i$ is purchased, product $j$ is frequently also purchased \cite{Leskovec07}. The other dataset has 1,696,415 vertices, representing nodes on the Internet, taken from autonomous system traceroutes in 2005 \cite{LKFGraphsOverTime}. The edges in this graph are undirected and represent communication links between nodes. In both cases, the 150 eigenvectors corresponding to the largest positive eigenvalues of the residuals matrices were computed, and subgraphs were analyzed that align with eigenvectors with small $L_1$ norms.

\begin{figure}
\begin{centering}
\includegraphics[width=1.7in]{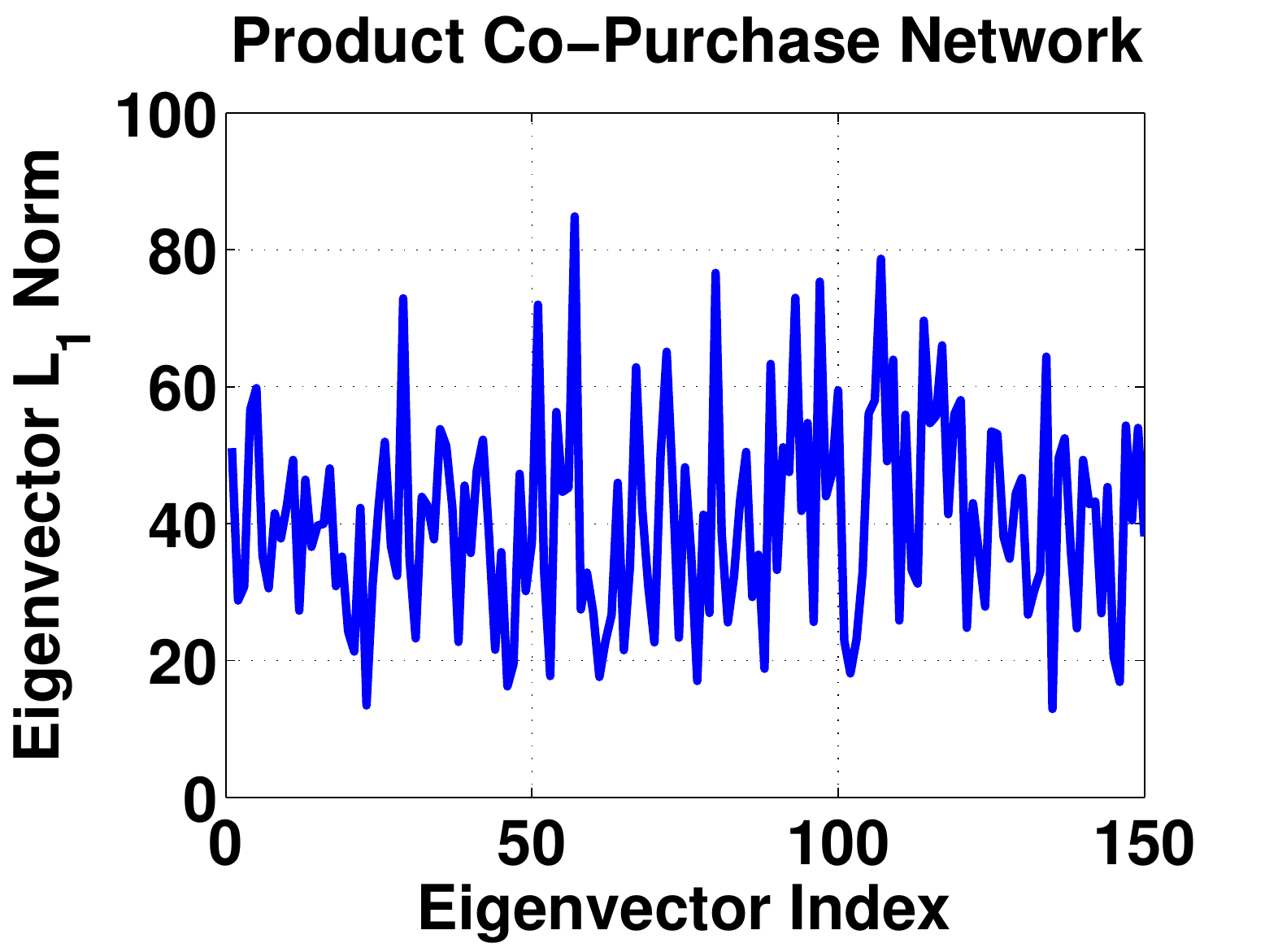}\ \ 
\includegraphics[width=1.7in]{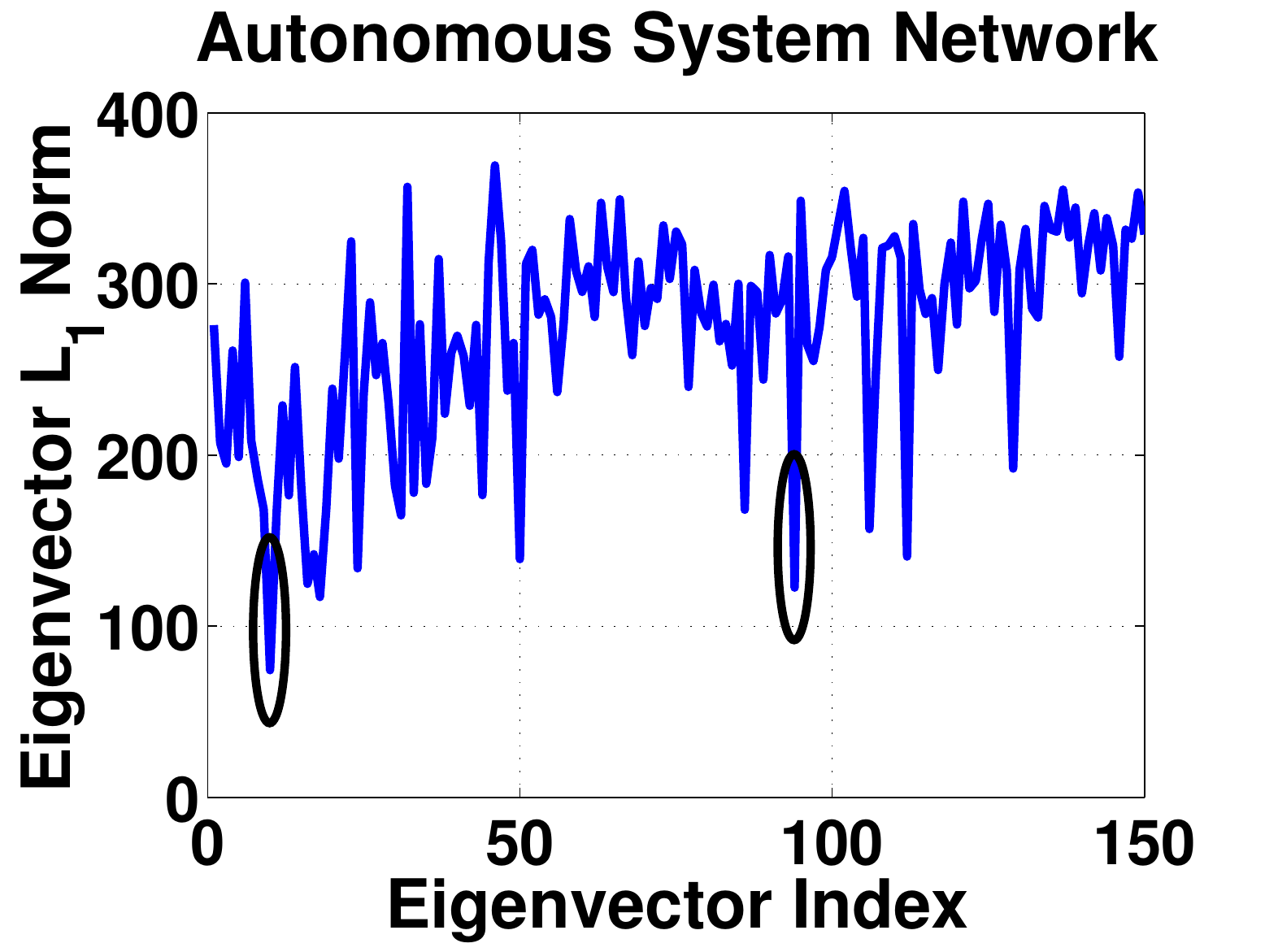}
\caption{Eigenvector $L_1$ norms in application datasets: an amazon.com product co-purchase network (left) and an autonomous system network (right). }
\label{fig:L1}
\end{centering}
\end{figure}

In the amazon.com co-purchase network, edges are directed, and each vertex has at most 5 outward edges. We use the symmetrized modularity matrix introduced in \cite{DirectedModularity} as a residuals matrix. As shown on the left in Fig. \ref{fig:L1}, many of the eigenvectors have small $L_1$ norms, due to frequent co-purchase of small, relatively isolated sets of products. We consider the 2 smallest $L_1$ norms, corresponding to the 23rd and 135th largest eigenvectors. 
 These eigenvectors are concentrated, respectively, on a 53-vertex subgraph with all possible internal edges (265) and a 44-vertex subgraph with 215 of its 220 possible internal edges. Neither subgraph has any outgoing edges, and both have fewer than 20 incoming edges. To compare this to the graph as a whole, we took 1 million samples of comparable size by performing random walks on the graph. Of all 53-vertex samples, only 609 have average internal degree greater than 4.5, and of those, none has fewer than 20 external edges. Similarly, among the random samples with 44 vertices, 108 have average internal degree greater than 4.4 and fewer than 40 external edges. Each of these 108 samples, however, is primarily outside of the 150-dimensional space spanned by the computed eigenvectors---an indicator vector for the sample vertices in each case is nearly in the null space of the matrix of eigenvectors. Thus, both of these subgraphs are anomalous with respect to random samples of similar size, when considering portions of the graph that are well-represented in the computed subspace.

The eigenvector $L_1$ norms in the autonomous system graph generally follow a trend, getting larger as the eigenvalues get smaller (indices increasing). The two vectors highlighted in the figure--the 10th and 94th--were considered for further investigation, since they have the largest local deviations. The 10th eigenvector is aligned with a 70-vertex subgraph with over 99\% of its possible edges, and the 94th eigenvector is aligned with a 28-vertex subgraph with over 81\% of its possible edges. These subgraphs consisted of primarily high-degree vertices, with average external degrees of about 957 and 577 for the 70- and 28-vertex subgraphs, respectively. We took 1 million random samples from among the subgraph of vertices with degree greater than 500, with sizes commensurate with the number of vertices in the subgraph within the high-degree vertex set (68 of 70 and 17 of 28). Among the three 68-vertex samples with density greater than 80\%, all share at least 55 vertices with the detected subgraph. Of the 17-vertex samples, 713 are at least 75\% dense and have fewer than 16,000 external edges (the 17-vertex subset is 93\% dense and has about 12,500 external edges). Of these 713 samples, all are significantly aligned with eigenvectors 10 and 18, both of which also have extremely small $L_1$ norms as shown in the figure. Thus, the only subgraphs among the samples with similar densities and external degrees are detectable through analysis of eigenvector $L_1$ norms.

\section{Conclusion}
\label{sec:conclusion}
In this paper, we present a spectral framework for the uncued detection of small anomalous signals within large, noisy background graphs. This framework is based on analysis of graph residuals in their principal eigenspace. We propose the spectral norm as a power metric, and several algorithms are outlined, with varying degrees of complexity. In simulation, we demonstrate the utility of the algorithms for detection and identification of two foregrounds within three background models, with the more computationally complex methods providing better detection performance. In two real networks, subgraphs detected via one of the algorithms are shown to be anomalous with respect to random samples of the background.

The framework presented in this paper demonstrates the utility of considering the anomalous subgraph detection problem in a signal processing context. There are myriad avenues of investigation from this point. Recent work has focused on extending this framework to time-varying graphs \cite{SSP2011, SSP2012} 
 and attributed graphs \cite{ISI2013}. Non-spectral statistics have also been of interest, in particular for detecting anomalously sparse (rather than anomalously dense) subgraphs \cite{ICASSP2012_Lauren}, though this complicates the analysis since embedding the signal involves subtracting edges rather than adding them. Another interesting area is detection using supervised learning based on subgraph features, as in \cite{Pan13}. Performance bounds in spectral detection of cliques and communities have recently been studied \cite{NadakuditiPlantedClique, NadakuditiNewmanCommDet}, as have computational limits of detection \cite{SPCAbound, tradeoffs}. Also, while the presented framework relies on analysis of residuals, considering \emph{normalized} residuals may improve detection for subgraphs where the edges are extremely unlikely \cite{RandomGraphSpectra, NadakuditiNewmanCL}. This analysis, however, may be intractable for more complicated graph models, since it requires normalizing each observed vertex pair and may not allow the computational tricks mentioned in Section \ref{subsec:overview}. As the detection of anomalous behavior in relational datasets continues to be a problem of interest, the field of signal processing for graphs will continue to pose a rich set of challenges for the research community.

\appendices
\section{Proof of Theorem \ref{thm:ERinER}}
\label{app:likelihoodProof}
Under $\calH_0$, the hypothesis that the observed graph was generated by an Erd\H{o}s--R\'{e}nyi process, the likelihood of the observed graph is given by
\begin{equation}
\calL(G; \calH_0, p)=p^{|E|}(1-p)^{\binom{N}{2}-|E|}.
\end{equation}
Under the alternative hypothesis, an $N_S$-vertex subset was selected uniformly at random to serve as the subgraph. Suppose that $V_S\subset V$, $|V_S|=N_S$, was chosen as the subset. Each pair of vertices within $V_S$ still has probability $p$ of sharing an edge due to background activity. If there is no edge in the background, however, an edge will be added with probability $p_S$. Thus, the probability of an edge occurring between a given pair of vertices both in $V_S$ is 
\begin{equation}
\hat{p}=p+(1-p)p_S=p+p_s-p\cdot p_S.
\end{equation}
All other vertex pairs still have probability $p$ of sharing an edge. Therefore, we have
\begin{align}
\calL(G; \calH_1, p, V_S, p_S)=&\ \hat{p}^{|E_S|}(1-\hat{p})^{\binom{N_S}{2}-|E_S|}\\
 & \cdot p^{|E|-|E_S|}(1-p)^{\binom{N}{2}-|E|-\binom{N_S}{2}+|E_S|}.\nonumber
\end{align}
Note that $\binom{N}{2}-|E|-\binom{N_S}{2}+|E_S|$ is the number of ``non-edges'' that are not within the subgraph vertices. Since only one vertex subset is chosen for the signal embedding, the likelihood of $G$ under the alternative hypothesis is
\begin{equation}
\sum_{V_S\subset V, |V_S|=N_S}{\calL(G; \calH_1, p, V_S, p_S)\operatorname{Pr}\left[V_S\textrm{ is chosen}\right]}.
\end{equation}
Each of the $\binom{N}{N_S}$ possible subsets is equally likely, so the likelihood ratio is 
\begin{equation}
\frac{\binom{N}{N_S}^{-1}\sum_{V_S\subset V, |V_S|=N_S}{\calL(G; \calH_1, p, V_S, p_S)}}{\calL(G; \calH_0, p)}
\end{equation}
or, equivalently
\begin{equation}
\binom{N}{N_S}^{-1}\sum_{V_S\subset V, |V_S|=N_S}{\frac{\calL(G; \calH_1, p, V_S, p_S)}{\calL(G; \calH_0, p)}}.\label{eqn:ratioSummand}
\end{equation}
The ratio in (\ref{eqn:ratioSummand}) can be further simplified as 
\begin{equation}
\frac{\calL(G; \calH_1, p, V_S, p_S)}{\calL(G; \calH_0, p)}=\left(\frac{1-\hat{p}}{1-p}\right)^{\binom{N_S}{2}}\left[\frac{\hat{p}(1-p)}{p(1-\hat{p})}\right]^{|E_S|}.\label{eqn:simplified}
\end{equation}
Replacing the ratio in (\ref{eqn:ratioSummand}) with the expression in (\ref{eqn:simplified}), and moving the non-subgraph-dependent portion outside of the summation, yields the expression in (\ref{eqn:ERinERlikelihood}). This completes the proof.

\section{Proof of Theorem \ref{thm:SN}}
\label{app:SNProof}
Let $u_1$ be the (unit-normalized) principal eigenvector of $\Ahat$. Since $u$ is the eigenvector corresponding to the largest eigenvalue of $B+\Ahat$, we have 
\begin{equation}
u_1^T(B+\Ahat)u_1=\|\Ahat\|+u_1^TBu_1\leq u^T(B+\Ahat)u.
\end{equation}
Since $u_1$ only has nonzero entries in rows corresponding to subgraph vertices, we can bound this quantity below by $\|\Ahat\|-\|B_S\|$.

The vector $u$ can be decomposed as $u=u_S+u_B$, where the only nonzero components of $u_S$ correspond to the signal subgraph vertices and $u_B$ may only be nonzero in the rows corresponding to $V\setminus V_S$. 
Let $\delta=\|u_S\|_2^2$. Since $u$ has unit $L_2$ norm, and $u_S$ and $u_B$ are orthogonal, we have $0\leq \delta\leq 1$ and $\|u_B\|_2^2=1-\delta$. The largest eigenvalue of the residuals matrix is then given by
\begin{align}
u^T(\Ahat+B)u=&\ u_S^T\Ahat u_S+2u_S^T\Ahat u_B+u_B^T\Ahat u_B\label{eqn:SNexpanded}\\
&+u_S^TBu_S+2u_S^TBu_B+u_B^TB u_B\nonumber.
\end{align}
Both terms that include $\Ahat u_B$ are zero, since $\Ahat$ is only nonzero within the subgraph vertices. To get an upper bound for this quantity, we bound each term in (\ref{eqn:SNexpanded}), yielding
\begin{align}
u^T(\Ahat+B)u\leq&\  \delta\|\Ahat\|+\delta\|B_{S}\|\label{eqn:UB}\\
&+2\sqrt{\delta(1-\delta)}\|B_{SN}\|+(1-\delta)\|B_{N}\|\nonumber.
\end{align}

For convenience, let $\alpha=\|\Ahat\|+\|B_{S}\|-\|B_{N}\|$, $\beta=2\|B_{SN}\|$, and $\gamma=\|B_{N}\|+\|B_{S}\|-\|\Ahat\|$. Combining the upper bound in (\ref{eqn:UB}) with the lower bound yields
\begin{equation}
\alpha\delta+\beta\sqrt{\delta(1-\delta)}+\gamma\geq0\label{eqn:quad}
\end{equation}
We can verify that, for $\beta\geq0$ and $-\alpha\leq\gamma<0$, the expression of (\ref{eqn:quad}) will achieve equality at the lesser of the two roots of the parabola obtained by squaring  the expression. Therefore, (\ref{eqn:quad}) holds whenever
\begin{equation}
\delta \geq \frac{\beta^2-2\alpha\gamma-\sqrt{\beta^4-4\beta^2\gamma(\alpha+\gamma)}}{2(\alpha^2+\beta^2)}.\label{eqn:LB}
\end{equation}
Using the triangle inequality to remove the radical in (\ref{eqn:LB}) and substituting the matrix norms back into the equation yields the bound in (\ref{eqn:normBound}). 
This completes the proof.

\section{Concentration of Eigenvectors on Subgraph Vertices}
\label{app:L1}
Here we provide an example of an embedding on which a single eigenvector will be concentrated. Consider a subgraph $\Ahat$ that is regular, i.e., each vertex has the same degree $d_S$. Such a subgraph will have a spectral norm $\|\Ahat\|=d_S$, and the principal eigenvector will be a vector in which all components are the same. Let $x$ be a unit-normalized indicator vector for the subgraph, i.e., a vector where the $i$th component is $1/\sqrt{N_S}$ if $i$ corresponds to a subgraph vertex and is $0$ otherwise. Further consider $x^T(B+\Ahat)x$ and $x^T(B+\Ahat)^2x$. We have 
\begin{equation}
x^T(B+\Ahat)x=d_S+x^TBx=d_S+X,
\end{equation}
where
\begin{equation}
X=\frac{1}{N_S}\sum_{i,j\in V_S}{(a_{ij}-p_{ij})}
\end{equation}
is a random variable whose mean is $0$ and variance is less than $\frac{2}{N_S^2}\expect\left[\left|E\cap(V_S\times V_S)\right|\right]$, that is, the expected fraction of possible edges between the subgraph vertices that exist in the background. If the embedding occurs on vertices where the expected connectivity is low, then $X$ will likely be very small. We also have
\begin{align}
x^T(B+\Ahat)^2x=&d_S^2+2d_SX+x^TB^2x\\
=&d_S^2+2d_SX+Y.\nonumber
\end{align}
Note that $Y=\|Bx\|_2^2=x^TB^2x$, which can be rewritten as
\begin{align}
x^TB^2x=&\frac{1}{N_S}\sum_{i=1}^{N}{\left[\sum_{j\in V_S}{(a_{ij}-p_{ij})}\right]^2}\\
=&\frac{1}{N_S}\sum_{i=1}^{N}{\sum_{j,k\in V_S}{(a_{ij}a_{ik}-a_{ij}p_{ik}-a_{ik}p_{ij}+p_{ij}p_{ik})}}.\nonumber
\end{align}
For $j\neq k$, the expectation of the summand is 0. Considering only $j=k$, we have
\begin{align}
\expect\left[x^TB^2x\right]=&\frac{1}{N_S}\sum_{i=1}^{N}\sum_{j\in{V_S}}{p_{ij}(1-p_{ij})}\\
<&\frac{1}{N_S}\sum_{i=1}^{N}\sum_{j\in{V_S}}{p_{ij}},
\end{align}
where the upper bound is the average expected degree of the subgraph vertices before the embedding occurs. Again, if the subgraph is embedded on vertices with low expected degree, this quantity is likely to be small.

Let $U\Lambda U^T=B+\Ahat$ be the eigendecomposition of the residuals matrix, with $\lambda_i$ denoting the $i$th eigenvector ($\lambda_i\geq \lambda_j$ for $i<j$), and let $z=U^Tx$.
We have
\begin{align}
\left(x^T(B+\Ahat)x\right)^2=&\left(\sum_{i=1}^{N}{\lambda_iz_i^2}\right)^2\nonumber\\
=&(d_S+X)^2\label{eqn:lambdaSumSquared}\\ 
=&d_S^2+2d_SX+X^2\nonumber
\end{align}
and 
\begin{equation}
x^T(B+\Ahat)^2x=\sum_{i=1}^{N}{\lambda_i^2z_i^2}=d_S^2+2d_SX+Y.
\label{eqn:lambdaSquaredSum}
\end{equation}
If the quantities in (\ref{eqn:lambdaSumSquared}) and (\ref{eqn:lambdaSquaredSum}) were the same, then $x$ would be an eigenvector of $B+\Ahat$. Since their difference is very small (i.e., assuming $Y$ and $X^2$ are small, as they are in expectation), then $x$ may be highly correlated with a single eigenvector. That is, for some $i$, $z_i^2$ may be quite large, so that $x$ concentrates most of its magnitude on the $i$th eigenvector. Let $\lambda_m$ be the eigenvalue closest to $d_S+X$, and $\delta=d_S+X-\lambda_m$.
 Then we have
\begin{align}
d_S+X=&\lambda_m+\delta\label{eqn:lambdaSum}\\
 =& \sum_{i=1}^{m-1}{\lambda_iz_i^2} + \lambda_mz_m^2 + \sum_{i=m+1}^{N}{\lambda_iz_i^2}.\nonumber
\end{align}
For $i\neq m$, let $\Delta_i=\lambda_i-\lambda_m$. 
For convenience, define the following substitutions:
\begin{align}
a=&\sum_{i=1}^{m-1}{z_i^2}\\
b=&z_m^2\\
c=&\sum_{i=m+1}^{N}{z_i^2}\\
\eop=&\frac{\sum_{i=1}^{m-1}{\Delta_iz_i^2}}{a}\\
\eom=&\frac{\sum_{i=m+1}^{N}{\Delta_iz_i^2}}{c}.
\end{align}
Thus, $\lambda_m+\eop$ and $\lambda_m+\eom$ are convex combinations of the eigenvalues greater than $\lambda_m$ and less than $\lambda_m$, respectively. We can then express (\ref{eqn:lambdaSum}) as 
\begin{equation}
d_S+X=a(\lambda_m+\eop)+b\lambda_m+c(\lambda_m+\eom).\label{eqn:L}
\end{equation}
Similarly, letting 
\begin{equation}
\etp=\frac{\sum_{i=1}^{m-1}{\Delta_i^2z_i^2}}{a}
\end{equation}
and
\begin{equation}
\etm=\frac{\sum_{i=m+1}^{N}{\Delta_i^2z_i^2}}{c},
\end{equation}
(\ref{eqn:lambdaSquaredSum}) can be rewritten as 
\begin{align}
d_S^2+2Xd_S+Y=&\ a(\lambda_m^2+2\eop\lambda_m+\etp)+b\lambda_m^2\label{eqn:Lsquared}\\
&+c(\lambda_m^2+2\eom\lambda_m+\etm).\nonumber
\end{align}
Combining equations (\ref{eqn:L}) and (\ref{eqn:Lsquared}) and performing some algebraic manipulation yields the system of equations
\begin{align}
\delta&=a\eop+c\eom\\
\delta^2+Y-X^2&=a\etp+c\etm\\
1&=a+b+c,
\end{align}
which, solving for $b$, gives us
\small\begin{align}
b=&1-\frac{(\delta^2+Y-X^2)(\eop-\eom)-\delta(\etp-\etm)}{\eop\etm-\eom\etp}\nonumber\\
>&1-\frac{\delta^2+Y-X^2}{\min(\etp, \etm)}-\frac{\left|\delta\right|}{\min(\eop, -\eom)}\label{eqn:bBound}.
\end{align}\normalsize
If the eigenvalues around $\lambda_m$ are spread far apart, then $\eop$, $-\eom$, $\etp$ and $\etm$ will be relatively large, the fractions in (\ref{eqn:bBound}) will be small, and $x$ will be heavily concentrated on a single eigenvector. This is supported by the empirical results in Section \ref{sec:results}, where embedding clusters onto vertices with low expected degree yields separation in a single eigenvector.

\section{Change in Modularity Due to Subgraph Embedding}
\label{app:deltaK}
When using observed degree to estimate expected degree, the difference in the expected value terms caused by the signal is as follows. If no embedding occurs, the estimated expected value is $\|k\|_1^{-1}kk^T$, where $k$ is the observed degree vector resulting from the background noise. If an anomalous subgraph is embedded into the background, the degree vector is changed by $\khat=\Ahat\ones$. Since  $\Ahat$ consists of only edges within the subgraph that do not appear due to noise, the degree vector after embedding is $k+\khat$, and the volume is $\|k+\khat\|_1=\|k\|_1+\|\khat\|_1$. Thus, the difference between the modularity matrix with estimated expected degrees under $\calH_0$ and $\calH_1$ is 
\begin{align}
\DelK&=\frac{kk^T}{\|k\|_1}-\frac{\left(k+\khat\right)\left(k+\khat\right)^T}{\|k\|_1+\|\khat\|_1}\label{eqn:delK}\\
&=\frac{\|\khat\|_1kk^T-\|k\|_1\left(k\khat^T+\khat k^T+\khat\khat^T\right)}{\|k\|_1\left(\|k\|_1+\|\khat\|_1\right)}.\nonumber
\end{align}
To bound the strength of $\DelK$, we will bound the spectral norm of each summand in the numerator of (\ref{eqn:delK}) and ignore the $\|\khat\|_1$ in the denominator, yielding
\begin{equation}
\|\DelK\|\leq\frac{\|\khat\|_1\|k\|_2^2+2\|k\|_1\|k\|_2\|\khat\|_2+\|k\|_1\|\khat\|_2^2}{\|k\|_1^2}.
\end{equation}
To show that the strength of this quantity will grow more slowly than the signal strength, given certain conditions, we will show that $\|\DelK\|/\|\Ahat\|$ is $o(1)$, i.e., that
\begin{equation}
\frac{\left(\|\khat\|_1\|k\|_2^2+2\|k\|_1\|k\|_2\|\khat\|_2+\|k\|_1\|\khat\|_2^2\right)}{\|k\|_1^2\|\Ahat\|}\rightarrow 0.
\end{equation}
Since $N_S\ll N$, we will ignore the $\|k\|_1\|\khat\|_2^2$ term, as the other terms will dominate it. Thus, we must bound 
\begin{align}
\frac{\|\khat\|_1\|k_2^2+2\|k\|_1\|k\|_2\|\khat\|_2}{\|k\|_1^2\|\Ahat\|}=&\frac{2\|k\|_2}{\|k\|_1}\cdot\frac{\|\khat\|_2}{\|\Ahat\|}\label{eqn:toBound}\\
&+\left(\frac{\|k\|_2}{\|k\|_1}\right)^2\frac{\|\khat\|_1}{\|\Ahat\|}.\nonumber
\end{align}

In many applications, the graphs of interest have degree sequences that follow a power law; i.e., the number of vertices with degree $i$ is approximately $\alpha i^{-\beta}$ for constants $\alpha ,\beta>0$. Using this model, we can analyze the ratio of $L_1$ and $L_2$ norms in graphs with a realistic growth pattern. Let $\kmax$ be the largest degree in the graph. Then the squares of the $L_1$ and $L_2$ norms of $k$ can be approximated as
\begin{equation}
\|k\|_1^2\approx\left(\sum_{i=1}^{\kmax}{i\cdot \alpha i^{-\beta}}\right)^2=\left(\sum_{i=1}^{\kmax}{\alpha i^{1-\beta}}\right)^2
\end{equation}
and
\begin{equation}
\|k\|_2^2\approx\sum_{i=1}^{\kmax}{i^2\cdot \alpha i^{-\beta}}=\sum_{i=1}^{\kmax}{\alpha i^{2-\beta}},
\end{equation}
respectively. Their ratio is then approximated, assuming $\beta$ does not exactly equal 1 or 2, as
\begin{align}
\left(\frac{\|k\|_2}{\|k\|_1}\right)^2\approx&\frac{\sum_{i=1}^{\kmax}{\alpha i^{2-\beta}}}{\left(\sum_{i=1}^{\kmax}{\alpha i^{1-\beta}}\right)^2}\nonumber\\
<&\frac{1}{\alpha} \frac{\int_1^{\kmax+1}{x^{2-\beta}dx}}{\left(\int_2^{\kmax}{x^{1-\beta}dx}\right)^2}\label{eqn:ratio}\\
=&\frac{1}{\alpha} \frac{\frac{1}{3-\beta}[(\kmax+1)^{3-\beta}-1]}{\left(\frac{1}{2-\beta}[\kmax^{2-\beta}-2^{2-\beta}]\right)^2}\nonumber\\
=&\frac{(2-\beta)^2}{\alpha (3-\beta)}\frac{(\kmax+1)^{3-\beta}-1}{\kmax^{4-2\beta}-2(2\kmax)^{2-\beta}+2^{4-2\beta}}.\nonumber
\end{align}
In practice, $\beta$ is typically greater than 1 and less than 3 (see, e.g., \cite{Faloutsos3}), so the constant $(2-\beta)^2/(3-\beta)$ will be positive. As $\kmax$ increases, the ratio on the right will tend to $\kmax^{\beta-1}$. If we let the maximum degree increase, however, $\alpha $ should be allowed to increase as well, since this controls the number of vertices with a given degree. Assume $\kmax$ is a degree that will probably not occur in the graph. Specifically, for a small, constant threshold $t$, let $\kmax=\inf\left\{i\arrowvert \alpha i^{-\beta}<t\right\}.$ Since this means that
\begin{equation}
\alpha(\kmax-1)^{-\beta}\geq t,\label{eqn:alphaBound}
\end{equation}
we have
\begin{equation}
\frac{1}{\alpha}{\kmax^{\beta-1}}\leq\frac{1}{\alpha}(\sqrt[\beta]{\alpha/t}+1)^{\beta-1}=O\left(\alpha^{-1/\beta}\right).\label{eqn:maxBound}
\end{equation}
Using the approximation in (\ref{eqn:ratio}), the ratio of the $L_2$ and $L_1$ norms of $k$ is approximately $O\left(1/\sqrt[2\beta]{\alpha}\right)$.

To bound the term dependent on the subgraph, we have
\begin{equation}
\frac{\|\khat\|_2}{\|\Ahat\|}=\frac{\sqrt{\ones^T\Ahat^2\ones}}{\|\Ahat\|}\leq\frac{\sqrt{N_S\|\Ahat\|^2}}{\|\Ahat\|}=\sqrt{N_S}.\label{eqn:normRatio}
\end{equation}
This upper bound can be achieved if the subgraph is a clique or a star. Noting that $\|\khat\|_1\leq\sqrt{N_S}\|\khat\|_2$, we substitute (\ref{eqn:maxBound}) and (\ref{eqn:normRatio}) into (\ref{eqn:toBound}) to obtain 
\begin{align}
\frac{\|\DelK\|}{\|\Ahat\|}\approx&O\left(\sqrt{N_S/\sqrt[\beta]{\alpha}}+N_S/\sqrt[\beta]{\alpha}\right)\\
=&O\left(N_S/\sqrt[\beta]{\alpha}\right),\nonumber
\end{align}
meaning that $\|\DelK\|$ is $o(\|\Ahat\|)$ if $N_S$ is $o(\sqrt[\beta]{\alpha})$. Using (\ref{eqn:alphaBound}) as a lower bound for $\alpha$, this implies that $\|\DelK\|/\|\Ahat\|$ will vanish as the graph grows if $N_S$ grows more slowly than $\kmax$.

\section*{Acknowledgment}
The authors would like to thank Dr. B. Johnson and the Lincoln Laboratory Technology Office for supporting this work, and R. Bond, Dr. J. Ward, and D. Martinez for their managerial support. We would also like to thank N. Singh, for his early work on the method in Section \ref{subsec:SPCA}. Finally, we would like to thank Dr. R. S. Caceres, Dr. R. J. Crouser, Prof. A. O. Hero III, Dr. S. Kelley, Dr. A. Reuther, Dr. M. C. Schmidt, Dr. M. M. Wolf, and the anonymous referees for many helpful comments on this paper.

\ifCLASSOPTIONcaptionsoff
  \newpage
\fi



\bibliographystyle{IEEEtran}
\bibliography{FullBib}

\begin{thebibliography}{10}
\providecommand{\url}[1]{#1}
\csname url@samestyle\endcsname
\providecommand{\newblock}{\relax}
\providecommand{\bibinfo}[2]{#2}
\providecommand{\BIBentrySTDinterwordspacing}{\spaceskip=0pt\relax}
\providecommand{\BIBentryALTinterwordstretchfactor}{4}
\providecommand{\BIBentryALTinterwordspacing}{\spaceskip=\fontdimen2\font plus
\BIBentryALTinterwordstretchfactor\fontdimen3\font minus
  \fontdimen4\font\relax}
\providecommand{\BIBforeignlanguage}[2]{{%
\expandafter\ifx\csname l@#1\endcsname\relax
\typeout{** WARNING: IEEEtran.bst: No hyphenation pattern has been}%
\typeout{** loaded for the language `#1'. Using the pattern for}%
\typeout{** the default language instead.}%
\else
\language=\csname l@#1\endcsname
\fi
#2}}
\providecommand{\BIBdecl}{\relax}
\BIBdecl

\bibitem{yeast}
D.~Bu, Y.~Zhao, L.~Cai, H.~Xue, X.~Zhu, H.~Lu, J.~Zhang, S.~Sun, L.~Ling,
  N.~Zhang, G.~Li, and R.~Chen, ``Topological structure analysis of the
  protein--protein interaction network in budding yeast,'' \emph{Nucleic Acids
  Research}, vol.~31, no.~9, pp. 2443--2450, 2003.

\bibitem{eigenspaceAnomalyDetection}
T.~Id\'{e} and H.~Kashima, ``Eigenspace-based anomaly detection in computer
  systems,'' in \emph{Proc. ACM Int. Conf. Knowledge Discovery and Data
  Mining}, 2004, pp. 440--449.

\bibitem{newman06}
M.~E.~J. Newman, ``Finding community structure in networks using the
  eigenvectors of matrices,'' \emph{Phys. Rev. E}, vol.~74, no.~3, 2006.

\bibitem{Xu14b}
K.~S. Xu and A.~O. {Hero III}, ``Dynamic stochastic blockmodels for
  time-evolving social networks,'' \emph{IEEE J. Sel. Topics Signal Process.},
  2014, to appear, preprint available: arXiv:1403.0921.

\bibitem{authoritativeSources}
J.~M. Kleinberg, ``Authoritative sources in a hyperlinked environment,''
  \emph{J. ACM}, vol.~46, no.~5, pp. 604--632, September 1999.

\bibitem{changeDetectionSAR}
K.~Chen, C.~Huo, Z.~Zhou, and H.~Lu, ``Unsupervised change detection in {SAR}
  image using graph cuts,'' in \emph{IEEE Int. Geoscience and Remote Sensing
  Symp.}, vol.~3, July 2008, pp. 1162--1165.

\bibitem{DSPonGraphs}
A.~Sandryhaila and J.~M.~F. Moura, ``Discrete signal processing on graphs,''
  \emph{IEEE Trans. Signal Process.}, vol.~61, pp. 1644--1656, April 2013.

\bibitem{SPoG}
D.~I. Shuman, S.~K. Narang, P.~Frossard, A.~Ortega, and P.~Vandergheynst, ``The
  emerging field of signal processing on graphs: Extending high-dimensional
  data analysis to networks and other irregular domains,'' \emph{IEEE Signal
  Processing Mag.}, vol.~30, pp. 83--98, May 2013.

\bibitem{randomTerroristTransactions}
T.~L. Mifflin, C.~Boner, G.~A. Godfrey, and J.~Skokan, ``A random graph model
  for terrorist transactions,'' in \emph{Proc. IEEE Aerospace Conf.}, 2004, pp.
  3258--3264.

\bibitem{Alon1998}
N.~Alon, M.~Krivelevich, and B.~Sudakov, ``Finding a large hidden clique in a
  random graph,'' in \emph{Proc. ACM-SIAM Symp. Discrete Algorithms}, 1998, pp.
  594--598.

\bibitem{NadakuditiPlantedClique}
R.~R. Nadakuditi, ``On hard limits of eigen-analysis based planted clique
  detection,'' in \emph{Proc. IEEE Statistical Signal Process. Workshop}, 2012,
  pp. 129--132.

\bibitem{AriasCastro13}
E.~Arias-Castro and N.~Verzelen, ``Community detection in random networks,''
  2013, preprint: arXiv.org:1302.7099.

\bibitem{ACSparse2013}
N.~Verzelen and E.~Arias-Castro, ``Community detection in sparse random
  networks,'' 2013, preprint: arXiv:1308.2955.

\bibitem{anomDet07}
W.~Eberle and L.~Holder, ``Anomaly detection in data represented as graphs,''
  \emph{Intelligent Data Analysis}, vol.~11, no.~6, pp. 663--689, December
  2007.

\bibitem{Skillicorn07}
D.~B. Skillicorn, ``Detecting anomalies in graphs,'' in \emph{Proc. IEEE
  Intelligence and Security Informatics}, 2007, pp. 209--216.

\bibitem{SmithICASSP2012}
S.~T. Smith, S.~Philips, and E.~K. Kao, ``Harmonic space-time threat
  propagation for graph detection,'' in \emph{Proc. IEEE Int. Conf. Acoust.,
  Speech and Signal Process.}, 2012, pp. 3933--3936.

\bibitem{SmithNetworkDetection2}
S.~T. Smith, K.~D. Senne, S.~Philips, E.~K. Kao, G.~Bernstein, and S.~Philips,
  ``Bayesian discovery of threat networks,'' \emph{IEEE Trans. Signal
  Process.}, 2014, to be published.

\bibitem{Coppersmith12}
G.~A. Coppersmith and C.~E. Priebe, ``Vertex nomination via content and
  context,'' 2012, preprint: arXiv.org:1201.4118v1.

\bibitem{FortunatoSurvey}
S.~Fortunato, ``Community detection in graphs,'' \emph{Physics Reports}, vol.
  486, pp. 75--174, February 2010.

\bibitem{NewmanGirvan04}
M.~E.~J. Newman and M.~Girvan, ``Finding and evaluating community structure in
  networks,'' \emph{Phys. Rev. E}, vol.~69, no.~2, 2004.

\bibitem{scanStats}
C.~E. Priebe, J.~M. Conroy, D.~J. Marchette, and Y.~Park, ``Scan statistics on
  {E}nron graphs,'' \emph{Computational \& Mathematical Organization Theory},
  vol.~11, no.~3, pp. 229--247, 2005.

\bibitem{CLR}
T.~H. Cormen, C.~E. Leiserson, and R.~L. Rivest, \emph{Introduction to
  Algorithms}.\hskip 1em plus 0.5em minus 0.4em\relax MIT Press, 1990.

\bibitem{RuanZhangSpectral}
J.~Ruan and W.~Zhang, ``An efficient spectral algorithm for network community
  discovery and its applications to biological and social networks,'' in
  \emph{Proc. IEEE Int. Conf. Data Mining}, 2007, pp. 643--648.

\bibitem{WhiteSmythClustering}
S.~White and P.~Smyth, ``A spectral clustering approach to finding communities
  in graphs,'' in \emph{Proc. SIAM Int. Conf. Data Mining}, 2005.

\bibitem{algebraicModularity}
D.~Fasino and F.~Tudisco, ``An algebraic analysis of the graph modularity,''
  2013, preprint: arXiv:1310.3031.

\bibitem{DingKolaczyk13}
Q.~Ding and E.~D. Kolaczyk, ``A compressed {PCA} subspace method for anomaly
  detection in high-dimensional data,'' \emph{IEEE Trans. Inf. Theory},
  vol.~59, November 2013.

\bibitem{anomEigenCompression}
S.~Hirose, K.~Yamanishi, T.~Nakata, and R.~Fujimaki, ``Network anomaly
  detection based on eigen equation compression,'' in \emph{Proc. ACM Int.
  Conf. Knowledge Discovery and Data Mining}, 2009, pp. 1185--1193.

\bibitem{ChungSGTBook}
F.~R.~K. Chung, \emph{Spectral Graph Theory}.\hskip 1em plus 0.5em minus
  0.4em\relax American Mathematical Society, 1997.

\bibitem{RandomDotProduct}
S.~J. Young and E.~R. Scheinerman, ``Random dot product graph models for social
  networks,'' in \emph{Algorithms and Models for the Web-Graph}, ser. LNCS,
  A.~Bonato and F.~R.~K. Chung, Eds.\hskip 1em plus 0.5em minus 0.4em\relax
  Springer, 2007, vol. 4863, pp. 138--149.

\bibitem{RandomGraphSpectra}
F.~Chung, L.~Lu, and V.~Vu, ``The spectra of random graphs with given expected
  degrees,'' \emph{Proc. Nat. Acad. Sci. USA}, vol. 100, no.~11, pp.
  6313--6318, 2003.

\bibitem{ICASSP2012}
B.~A. Miller, N.~Arcolano, M.~S. Beard, J.~Kepner, M.~C. Schmidt, N.~T. Bliss,
  and P.~J. Wolfe, ``A scalable signal processing architecture for massive
  graph analysis,'' in \emph{Proc. IEEE Int. Conf. Acoust., Speech and Signal
  Process.}, 2012, pp. 5329--5332.

\bibitem{PerryWolfe12}
P.~O. Perry and P.~J. Wolfe, ``Null models for network data,'' 2012, preprint:
  arXiv:1201.5871v1.

\bibitem{ChoiWolfeAiroldi11}
D.~S. Choi, P.~J. Wolfe, and E.~M. Airoldi, ``Stochastic blockmodels with
  growing number of classes,'' \emph{Biometrika}, vol.~99, no.~2, pp. 273--284,
  2012.

\bibitem{ICASSP2010}
B.~A. Miller, N.~T. Bliss, and P.~J. Wolfe, ``Toward signal processing theory
  for graphs and non-{E}uclidean data,'' in \emph{Proc. IEEE Int. Conf.
  Acoust., Speech and Signal Process.}, 2010, pp. 5414--5417.

\bibitem{NIPS2010}
------, ``Subgraph detection using eigenvector {L1} norms,'' in \emph{Advances
  in Neural Inform. Process. Syst. 23}, J.~Lafferty, C.~K.~I. Williams,
  J.~Shawe-Taylor, R.~Zemel, and A.~Culotta, Eds., 2010, pp. 1633--1641.

\bibitem{SSP2011_Navraj}
N.~Singh, B.~A. Miller, N.~T. Bliss, and P.~J. Wolfe, ``Anomalous subgraph
  detection via sparse principal component analysis,'' in \emph{Proc. IEEE
  Statistical Signal Process. Workshop}, 2011, pp. 485--488.

\bibitem{SPGLLJ}
B.~A. Miller, N.~T. Bliss, P.~J. Wolfe, and M.~S. Beard, ``Detection theory for
  graphs,'' \emph{Lincoln Laboratory J.}, vol.~20, no.~1, 2013.

\bibitem{Don06}
D.~Donoho, ``Compressed sensing,'' \emph{IEEE Trans. Inf. Theory}, vol.~52,
  no.~4, pp. 1289--1306, 2006.

\bibitem{Eigenspokes}
B.~A. Prakash, A.~Sridharan, M.~Seshadri, S.~Machiraju, and C.~Faloutsos,
  ``{EigenSpokes}: Surprising patterns and scalable community chipping in large
  graphs,'' in \emph{Advances in Knowledge Discovery and Data Mining}, ser.
  LNCS, M.~J. Zaki, J.~X. Yu, B.~Ravindran, and V.~Pudi, Eds.\hskip 1em plus
  0.5em minus 0.4em\relax Springer, 2010, vol. 6119, ch.~14, pp. 435--448.

\bibitem{WWLZKurtosis}
L.~Wu, X.~Wu, A.~Lu, and Z.-H. Zhou, ``A spectral approach to detecting subtle
  anomalies in graphs,'' \emph{J. Intelligent Inform. Syst.}, vol.~41, no.~2,
  pp. 313--337, 2013.

\bibitem{SDPforPCA}
A.~{d'Aspremont}, L.~E. Ghaoui, M.~I. Jordan, and G.~R.~G. Lanckriet, ``A
  direct formulation for sparse {PCA} using semidefinite programming,''
  \emph{SIAM Review}, vol.~49, no.~3, pp. 434--448, 2007.

\bibitem{IRLanczos}
R.~Lehoucq and D.~Sorensen, ``Implicitly restarted {Lanczos} method,'' in
  \emph{Templates for the Solution of Algebraic Eigenvalue Problems: A
  Practical Guide}, Z.~Bai, J.~Demmel, J.~Dongarra, A.~Ruhe, and H.~van~der
  Vorst, Eds.\hskip 1em plus 0.5em minus 0.4em\relax Philadelphia: SIAM, 2000,
  ch. 4.5.

\bibitem{ER}
P.~Erd\H{o}s and A.~R\'{e}nyi, ``On random graphs,'' \emph{Publicationes
  Mathematicae Debrecen}, vol.~6, pp. 290--297, 1959.

\bibitem{BAPA99}
A.~Barab\'{a}si and R.~Albert, ``Emergence of scaling in random networks,''
  \emph{Science}, vol. 286, no. 5439, pp. 509--512, 1999.

\bibitem{ICASSP2012_Nick}
N.~Arcolano, K.~Ni, B.~A. Miller, N.~T. Bliss, and P.~J. Wolfe, ``Moments of
  parameter estimates for {C}hung--{L}u random graph models,'' in \emph{Proc.
  IEEE Int. Conf. Acoust., Speech and Signal Process.}, 2012, pp. 3961--3964.

\bibitem{RMAT}
D.~Chakrabarti, Y.~Zhan, and C.~Faloutsos, ``{R-MAT}: A recursive model for
  graph mining,'' in \emph{Proc. SIAM Int. Conf. Data Mining}, 2004, pp.
  442--446.

\bibitem{DSPCA}
\BIBentryALTinterwordspacing
R.~Luss, A.~{d'Aspremont}, and L.~E. Ghaoui, ``{DSPCA}: Sparse {PCA} using
  semidefinite programming,'' December 2008, version 0.6. [Online]. Available:
  \url{http://www.di.ens.fr/$\sim$aspremon/DSPCA.html}
\BIBentrySTDinterwordspacing

\bibitem{Leskovec07}
J.~Leskovec, L.~A. Adamic, and B.~A. Huberman, ``The dynamics of viral
  marketing,'' \emph{ACM Trans. Web}, vol.~1, pp. 1--39, May 2007.

\bibitem{LKFGraphsOverTime}
J.~Leskovec, J.~Kleinberg, and C.~Faloutsos, ``Graphs over time:
  {Densification} laws, shinking diameters and possible explanations,'' in
  \emph{Proc. Int. Conf. Knowledge Discovery and Data Mining}, 2005, pp.
  177--187.

\bibitem{DirectedModularity}
E.~A. Leicht and M.~E.~J. Newman, ``Community structure in directed networks,''
  \emph{Phys. Rev. Lett.}, vol. 100, pp. 118\,703--(1--4), Mar 2008.

\bibitem{SSP2011}
B.~A. Miller, M.~S. Beard, and N.~T. Bliss, ``Matched filtering for subgraph
  detection in dynamic networks,'' in \emph{Proc. IEEE Statistical Signal
  Process. Workshop}, 2011, pp. 509--512.

\bibitem{SSP2012}
B.~A. Miller and N.~T. Bliss, ``Toward matched filter optimization for subgraph
  detection in dynamic networks,'' in \emph{Proc. IEEE Statistical Signal
  Process. Workshop}, 2012, pp. 113--116.

\bibitem{ISI2013}
B.~A. Miller, N.~Arcolano, and N.~T. Bliss, ``Efficient anomaly detection in
  dynamic, attributed graphs,'' in \emph{Proc. IEEE Intelligence and Security
  Informatics}, 2013, pp. 179--184.

\bibitem{ICASSP2012_Lauren}
B.~A. Miller, L.~H. Stephens, and N.~T. Bliss, ``Goodness-of-fit statistics for
  anomaly detection in {C}hung--{L}u random graphs,'' in \emph{Proc. IEEE Int.
  Conf. Acoust., Speech and Signal Process.}, 2012, pp. 3265--3268.

\bibitem{Pan13}
S.~Pan and X.~Zhu, ``Graph classification with imbalanced class distributions
  and noise,'' in \emph{Proc. Int. Joint Conf. Artificial Intell.}, 2013, pp.
  1586--1592.

\bibitem{NadakuditiNewmanCommDet}
R.~R. Nadakuditi and M.~E.~J. Newman, ``Graph spectra and the detectability of
  community structure in networks,'' \emph{Phys. Rev. Lett.}, vol. 108, no.~18,
  pp. 188\,701--1--5, 2012.

\bibitem{SPCAbound}
Q.~Berthet and P.~Rigollet, ``Complexity theoretic lower bounds for sparse
  principal component detection,'' in \emph{Conf. Learning Theory}, ser. JMLR
  W\&CP, S.~Shalev-Shwartz and I.~Steinwart, Eds., 2013, vol.~30, pp.
  1046--1066.

\bibitem{tradeoffs}
Y.~Chen and J.~Xu, ``Statistical-computational tradeoffs in planted problems
  and submatrix localization with a growing number of clusters and
  submatrices,'' 2014, preprint: arXiv:1402.1267.

\bibitem{NadakuditiNewmanCL}
R.~R. Nadakuditi and M.~E.~J. Newman, ``Spectra of random graphs with arbitrary
  expected degrees,'' \emph{Phys. Rev. E}, vol.~87, no.~1, pp. 012\,803--1--12,
  2013.

\bibitem{Faloutsos3}
M.~Faloutsos, P.~Faloutsos, and C.~Faloutsos, ``On power-law relationships of
  the {Internet} topology,'' in \emph{Proc. SIGCOMM}, 1999.

\end{thebibliography}
\end{document}